\documentclass[useAMS,usenatbib]{mn2e}
\usepackage{graphicx,color,oldgerm}
\usepackage{amssymb,amsmath}
\usepackage{hyperref}
\usepackage{deluxetable}
\newcommand{\MBH}{\ensuremath{M_{\mathrm{BH}}}}
\newcommand{\Msun}{\ensuremath{{\rm M}_{\odot}}}
\newcommand{\Mstar}{\ensuremath{M_{\ast}}}
\newcommand{\Nstar}{\ensuremath{N_{\ast}}}
\newcommand{\Rstar}{\ensuremath{R_{\ast}}}
\newcommand{\Rnb}{\ensuremath{R_{\rm NB}}}
\newcommand{\Npart}{\ensuremath{N_{\rm p}}}
\newcommand{\Mcl}{\ensuremath{M_{\rm cl}}}
\newcommand{\Rsun}{\ensuremath{{\rm R}_{\odot}}}

\newcommand{\kms}{\ensuremath{\mathrm{km\,s}^{-1}}}

\newcommand{\Vrel}{\ensuremath{V_{\rm rel}^{\infty}}}
\newcommand{\Vstar}{\ensuremath{V_\ast}}
\newcommand{\trh}{\ensuremath{t_{\rm rh}}}
\newcommand{\trc}{\ensuremath{t_{\rm rc}}}
\newcommand{\tcc}{\ensuremath{t_{\rm cc}}}
\newcommand{\tccrlx}{\ensuremath{\left.t_{\rm cc}\right|_{\rm rlx}}}
\newcommand{\tra}{\ensuremath{t_{\rm ra}}}
\newcommand{\tstar}{\ensuremath{t_\ast}}
\newcommand{\tcoll}{\ensuremath{t_{\rm coll}}}
\newcommand{\tcollc}{\ensuremath{t_{\rm coll,c}}}
\newcommand{\trlx}{\ensuremath{t_{\rm rlx}}}
\newcommand{\Rwand}{\ensuremath{R_{\rm wand}}}
\newcommand{\Rcore}{\ensuremath{R_{\rm core}}}

\newcommand{\tKH}{\ensuremath{t_{\rm KH}}}
\newcommand{\Rh}{\ensuremath{R_{\rm h}}}
\newcommand{\GCoulomb}{\ensuremath{\gamma_{\rm c}}}
\newcommand{\sigmaOneD}{\ensuremath{\sigma_{v}}}
\newcommand{\sigmaThreeD}{\ensuremath{\sigma_{3}}}
\newcommand{\MESSY}{ME(SSY)**2}

\newcommand{\rem}[1]{} 
\newcommand{\Comment}[3]{}

\newcommand{\Sim}[1]{\texttt{#1}}

\title[Runaway collisions]{%
Runaway collisions in young star clusters. II.~Numerical results}

\author[M. Freitag, M. A. G\"urkan and F. A. Rasio]{
Marc Freitag$^{1,2}$\thanks{Present address: Institute of Astronomy, University of Cambridge, Madingley Road, Cambridge CB3~0HA, UK. E-mail: freitag@ast.cam.ac.uk},
M.\ Atakan G\"urkan$^{2,3}$ 
and
Frederic A.\ Rasio$^2$
\\
$^1$Astronomisches Rechen-Institut, M\"onchhofstrasse 12-14,
D-69120 Heidelberg, Germany\\
$^2$Department of Physics and Astronomy, 
Northwestern University, Evanston, IL 60208, USA\\
$^3$Foundations 
Development, Sabanc\i\ University, 34956 \.Istanbul, Turkey
}

\begin{document}

\date{Accepted. Received; in original form}

\pagerange{\pageref{firstpage}--\pageref{lastpage}} \pubyear{2005}

\maketitle

\label{firstpage}

\begin{abstract}
We present a new study of the collisional runaway scenario to form an
intermediate-mass black hole (IMBH, $\MBH\gtrsim 100\,\Msun$) at the
centre of a young, compact stellar cluster. The first phase is the
formation of a very dense central core of massive stars
($\Mstar\simeq30-120\,\Msun$) through mass segregation and
gravothermal collapse. Previous work established the
conditions for this to happen before the massive stars evolve off the
main sequence (MS). In this and a companion paper, we investigate the
next stage by implementing direct collisions between stars. Using a
Monte Carlo stellar dynamics code, we follow the core collapse and
subsequent collisional phase in more than 100 models with varying
cluster mass, size, and initial concentration. Collisions are treated
either as ideal, ``sticky-sphere'' mergers or using realistic
prescriptions derived from 3-D hydrodynamics computations. In all cases
for which the core collapse happens in less than the MS lifetime of
massive stars ($\simeq 3\,$Myr), we obtain the growth of a single very
massive star (VMS, $\Mstar\simeq 400-4000\,\Msun$) through a runaway
sequence of mergers. Mass loss from collisions, even for velocity
dispersions as high as $\sigmaOneD\sim 1000\,\kms$, does not prevent the
runaway. The region of cluster parameter space leading to runaway is
even more extended than predicted in previous work because, in
clusters with $\sigmaOneD>300\,\kms$, collisions accelerate (and, in
extreme cases, drive) core collapse. Although the VMS grows rapidly to
$\gtrsim 1000\,\Msun$ in models exhibiting runaway, we cannot predict
accurately its final mass. This is because the termination of the
runaway process must eventually be determined by a complex interplay
between stellar dynamics, hydrodynamics, and the stellar evolution of
the VMS. In the vast majority of cases, we find that the time between
successive collisions becomes much shorter than the thermal
timescale of the VMS. Therefore, our assumption that all stars return
quickly to the MS after a collision must eventually break down for the
runaway product, and the stellar evolution of the VMS becomes very
uncertain. For the same reason, the final fate of the VMS, including
its possible collapse to an IMBH, remains unclear.
\Comment{}{}{Introduce companion paper?}
\end{abstract}


\section{INTRODUCTION}

In our first paper (\citealt*{FRB05}; hereafter Paper~I), we have
presented our numerical approach to the study of stellar collisions in
young, dense star clusters with a broad stellar mass spectrum. It is
based on the use of {\MESSY}, a Monte Carlo code to simulate the
long-term evolution of spherical clusters subject to relaxation,
collisions, stellar evolution and taking into account the possible
presence of a central massive object \citep{FB01a,FB02b}. Our main
motivation is to investigate in detail the runaway growth of a massive
object (``very massive star'', hereafter VMS) during core collapse
(\citealt{EbisuzakiEtAl01}; \citealt{PZMcM02};
\citealt*{GFR04}, hereafter GFR04). This provides a
natural path to the formation of an intermediate-mass black hole
(IMBH) in a dense star cluster, or a seed black hole (BH) in a
proto-galactic nucleus In Paper~I, we presented a number of test
calculations to validate our Monte Carlo code and compare its results
for simple idealized systems to those from $N$-body, Fokker-Planck and
gaseous-model codes. These comparisons established that we can
reliably follow the relaxation-driven evolution of clusters with a
broad mass function, all the way to core collapse. We also found good
agreement with models of dense clusters in which mergers between stars
create a mass spectrum starting from a single-mass population, thus
accelerating collapse and leading to collisional runaway \citep{QS90}.

Here we use {\MESSY} to perform a large number of calculations
covering broadly the parameter space of young star
clusters. Specifically, we vary systematically the number of stars (in
the range $10^5-10^8$, using up to $9\times 10^6$ particles) and the
size of the cluster (with half-mass radius values from $0.02$ to
$5\,$pc), and we consider systems with low and high initial
concentration (King parameter $W_0=3$ and $W_0=8$). Our work is guided
by the main finding of GFR04, namely that for clusters with realistic
initial mass functions (for which the ratio of the maximum stellar
mass to the average mass is larger than $\sim100$), mass segregation
drives the cluster to core collapse in a time not longer than {\em
15\% of the initial central relaxation time}. It is therefore expected
that in any cluster for which this relaxational core-collapse time is
shorter that the main-sequence (MS) lifetime of the most massive stars
($\Mstar\approx 100\,\Msun$), i.e., $\sim 3\,$Myr, these objects will
eventually collide with each other in the high-density core, likely in
a runaway fashion. As we show in Sec.~\ref{sec:simulations}, our
results confirm this expectation.

Our work on the subject was inspired by the investigations of 
Portegies Zwart and collaborators who revived the study of the runaway
scenario thanks to $N-$body simulations \citep{PZMMcMH99,
PZMcM02}. Although many key aspects of the process were already known
from older works (in particular
\citealt{SS66,Colgate67,Sanders70b,BR78,Vishniac78,Lee87,QS90}; see paper~I
for a review of the field), these recent $N-$body studies were seminal
in considering the effects of collisions in a cluster with a
realistically broad IMF. This demonstrated explicitly for the first
time how mass segregation can lead to a collisional phase by
concentrating massive stars in a small central volume and made it
clear that the post-MS evolution of the most massive stars initially
present can prevent the runaway phase by driving cluster
expansion. Furthermore, Portegies Zwart and collaborators showed that,
contrary to the somewhat unrealistic situation in single-mass clusters
\citep{Lee87,QS90}, dynamically formed binaries, far from preventing the 
runaway phase by halting the central concentration increase, promote
it, as many binary interactions lead to stellar mergers. As
noted in paper~I and suggested by $N-$body simulations
with higher particle numbers \citep{PZBHMM04}, in a broad-IMF cluster
with $\Nstar\gtrsim 10^6$, binaries probably cannot form through 3-body
interactions {\it before} the collisional phase is reached.

It is important to stress, however, that our MC simulations probe a
different regime from that explored by the direct $N-$body
approach. We consider systems with higher number of stars in the
central region (inside a few core radii). For this reason,
some of our most important results are only superficially similar to
the findings of Portegies Zwart et al. In particular, in such
large-$N$ clusters with a long initial central collision time, a
genuine core collapse, itself driven by mass segregation, appears to
be the condition for collisional runaway. The situation for systems
with a smaller number of stars is more complex because there may not
be enough massive stars in the central regions to drive such a core
collapse \citep{PZMMcMH99,PZMcM02,McMPZ04,PZBHMM04}.

Our paper is organised as follows. In Section~2, we present the
simulations we have performed and explain their results.  In
Section~3, we summarise our findings and discuss avenues for future
research in the field. Detailed presentations of the runaway scenario,
the physics at play, our numerical methods and test computations are
to be found in Paper~I.

\section{SIMULATIONS}
\label{sec:simulations}

\subsection{Initial conditions and units}
\label{subsec:inicond_units}

In this work, we consider the evolution of isolated spherical stellar
cluster with a broad, realistic initial mass function (IMF), subject
to two-body relaxation and stellar collisions.  When not stated
otherwise, we use a Salpeter IMF, for which the number,
$d\Nstar$, of stars with masses between $\Mstar$ and  $\Mstar+d\Mstar$ is given by
\begin{equation}
\frac{d{\Nstar}}{dM_\ast} \propto 
M_\ast^{-\alpha} \mbox{\ \ for \ \ } M_{\rm min}\le M_\ast \le M_{\rm max}
\end{equation}
with $\alpha=2.35$, $M_{\rm min}=0.2\,\Msun$ and $M_{\rm
max}=120\,\Msun$. For this stellar population, the average mass is
$\langle M_\ast \rangle \simeq 0.69\,\Msun$. There is no initial mass
segregation. To investigate the role of the initial concentration of
the cluster, we consider King models with $W_0=3$ or $W_0=8$
\citep[][Sec.~4.4]{BT87}. We do not enforce tidal truncation because
it was shown in GFR04 that this does not affect the central regions
which completely dominate the evolution.

We refer to GFR04 (Section~3 and Table 1) for detailed
explanations about the important physical parameters of such clusters
and units. In keep with the tradition, when not stated otherwise, we
are using the $N$-body unit system \citep{Henon71a} defined by $G=1$,
$M_{\rm cl}(0)=1$ (initial total cluster mass) and $U_{\rm
cl}(0)=-1/2$ (initial cluster potential energy). As time unit, we
prefer the ``Fokker-Planck'' time $T_{\rm FP}$ to the $N$-body unit
$T_{\rm NB}$ because the former is a relaxation time while the latter
is a dynamical time; they are related to each other by $T_{\rm FP} =
\Nstar/\ln(\GCoulomb\Nstar)\, T_{\rm NB}$ where $\Nstar$ is the total 
number of stars in the cluster. As explained in paper~I, we
conservatively use $\GCoulomb=0.01$ to determine relaxation times in
this work. For King models, the $N$-body length unit is close to the
half-mass radius, ${\Rnb}\simeq 1.19\,\Rh$ for $W_0=3$ and
${\Rnb}\simeq 1.15\,R_{\rm h}$ for $W_0=8$. The core radius
(Spitzer87, Eq.~1-34 and Paper~I) is $\Rcore \simeq 0.543\,{\Rnb}$ and
encompasses a fraction $0.238$ of the total mass for $W_0=3$; for
$W_0=8$, these values are $0.121\,{\Rnb}$ and $0.0531$, respectively.

Expressions for half-mass and local relaxation times are given in
GFR04 and Paper~I. We denote the initial values of the half-mass and
central relaxation times by $\trh(0)$ and $\trc(0)$, respectively.

Table~1 lists the initial conditions for all
runs performed in this study. $\Npart$ denotes the number of
particles used in the simulation. We also give in this table the
value $\tra$ of the time when runaway started for all runs in which a
VMS formed.

\subsection{Overview of the results of the standard set of simulations}
\label{subsec:standard_overview}

\begin{figure*}
  \centerline{\resizebox{0.9\hsize}{!}{%
          \includegraphics[bb=21 158 581 700,clip]{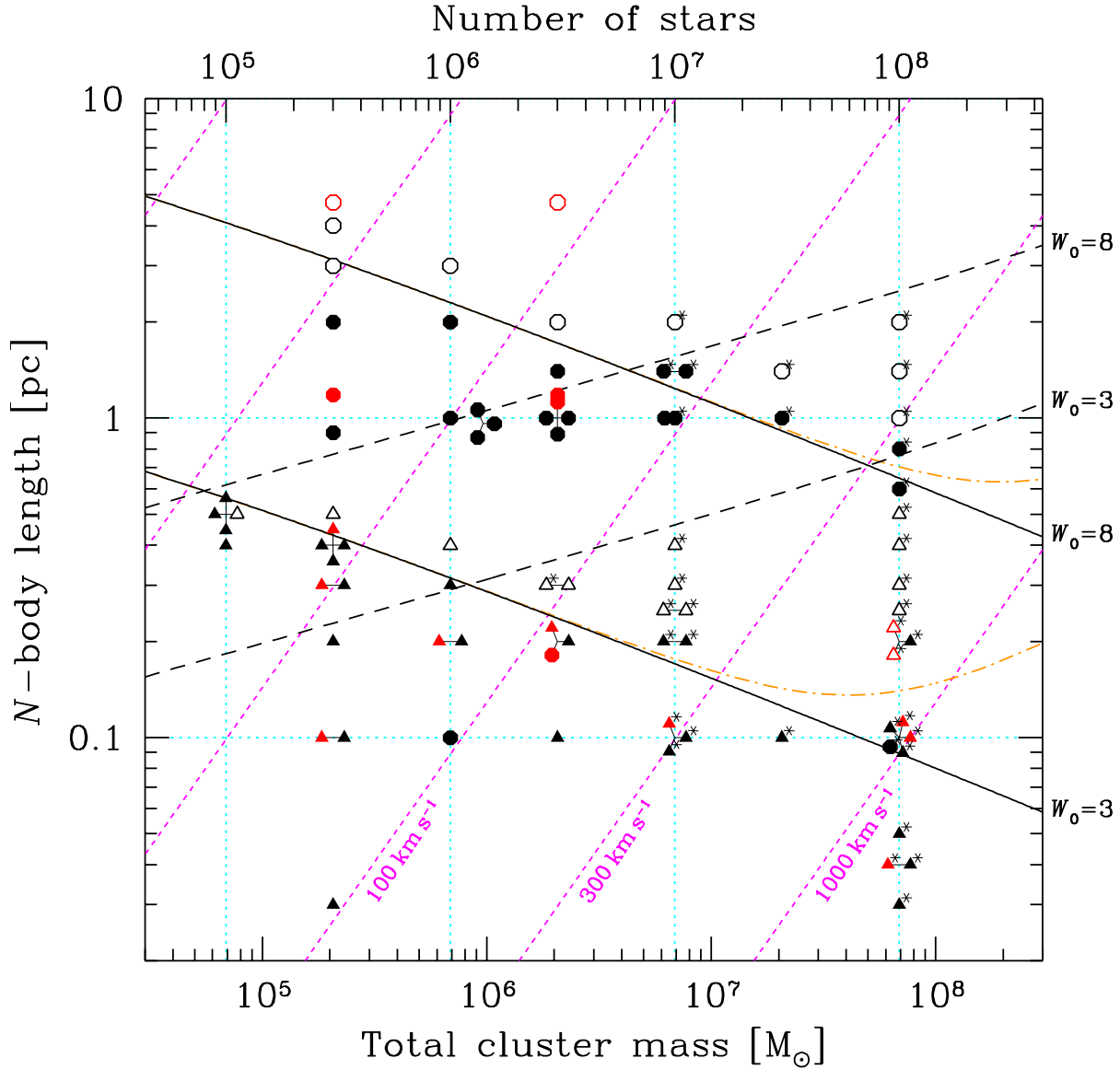}%
        }}
%
\caption{
  Diagram summarising the initial conditions and outcomes of the
  ``standard'' cluster simulations with collisions treated as perfect
  mergers or using the SPH prescription. Each point represents the
  cluster mass and (initial) $N$-body length unit (${\Rnb}$) for one
  simulation. Triangles correspond to $W_0=3$ King models, round dots
  to $W_0=8$. The standard $0.2-120\,\Msun$ Salpeter IMF was used in
  all cases. The solid diagonal lines (of negative slope) show the
  condition for the relaxation-driven core collapse time to be 3\,Myr,
  $\tccrlx = 0.12\,\trc(0) = 3\,$Myr for $W_0=8$ (top) and $W_0=3$
  (bottom). The long-dashed lines (of positive slope) show where the
  central collision time for a $120\,\Msun$ star (with any other star)
  is 3\,Myr. The dot-dashed curves indicate were the time for
  occurrence of runaway should be 3\,Myr, according to our
  parametrisation (see Eq.~\ref{eq:trunaway} and text). Short-dashed
  lines indicate approximately the 1-D central velocity
  dispersion. Solid symbols are for simulations which resulted in
  runaway formation of a VMS; open symbols are for cases in which
  stellar evolution interrupted core collapse before a VMS could
  grow. In some cases, several simulations with the same ${\Rnb}$ and
  ${\Nstar}$ (but different $W_0$, initial realisations of the
  cluster, random sequences for the MC algorithm, number of particles
  or collision prescription) were carried out. These cases correspond
  to symbols connected together with a thin line. Asterisks indicate
  simulations done with a number of particles smaller than the number
  of stars ($3\times 10^5$ or $10^6$ particles for $W_0=3$ cases,
  $3\times 10^6$ particles for $W_0=8$). In the colour on-line version
  of this diagram, simulations using the SPH prescriptions for
  collision outcome are indicated by red points. Among the three
  models with $W_0=3$, $\Nstar=10^8$ and $\Rnb=0.2\,$pc, one was
  computed with the sticky-sphere approximation and experienced
  runaway; the other two, making use of SPH prescriptions, missed the
  runaway phase.
\newline This is essentially the same plane as
  Fig.~1 of Paper~I except that ${\Rnb}$ is used instead
  of $R_{\rm h}$ and that the plotting domain is shifted to higher
  masses and smaller sizes to match the conditions that can be treated
  by the MC approach and may lead to runaway evolution.}
\label{fig:runaway_plane}
\end{figure*}

\begin{figure}
  \resizebox{\hsize}{!}{%
          \includegraphics[bb=30 164 564 688,clip]{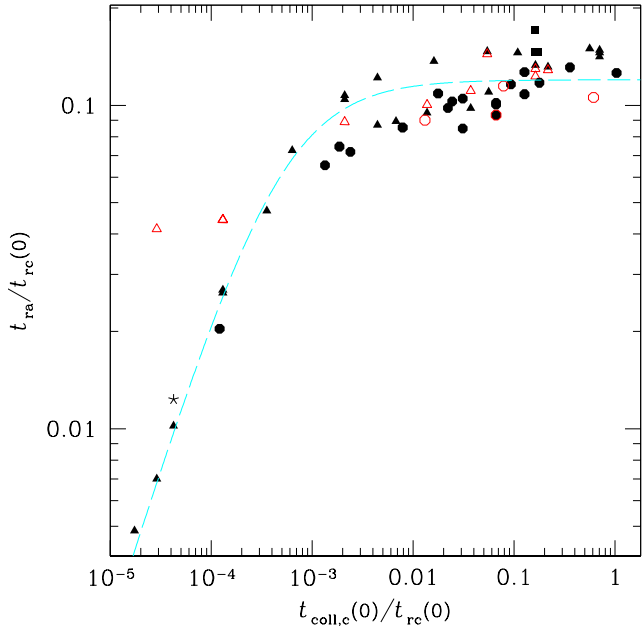}%
        }
\caption{
	Time of collisional runaway occurrence as a function of
	central initial collision time for $120\,\Msun$ stars
	(Eq.~\ref{eq:tcoll_ini}), for all simulations with standard
	Salpeter MF where runaway happened. Times are given in units
	of the initial central relaxation time. Triangles and circles
	are for $W_0=3$, $8$ models respectively. The squares are
	$W_0=9$, $12$ models for the cluster MGG-9 in M\,82 (see
	Sec.~\ref{subsec:MGG}). These simulations were carried out
	with a Kroupa MF extending from 0.1 to $100\,\Msun$. Solid
	symbols correspond to the ``sticky-sphere'' treatment of
	collisions, open symbols to the SPH prescriptions for
	collision outcomes. The dashed line is an ad-hoc
	parameterisation of the sticky-sphere results given in
	Eq.~\ref{eq:trunaway}. Note that for large collision times,
	one gets ${\tra}\simeq 0.12\,\trc(0)$, in good agreement with
	GFR04 for the relaxational core-collapse time but with a large
	dispersion in the result. The star corresponds to a simulation
	with no 2-body relaxation but the same initial conditions as
	those for the triangular point below it (see
	Fig.~\ref{fig:LagrRadHiColl}).}
\label{fig:trunaway}
\end{figure}

We first establish the conditions for runaway formation of a VMS
using the simplest and most favourable prescription for the outcome of
collisions, i.e., assuming that every time two stars come closer to
each other than the sum of their radii, they merge without any mass
loss. With such a clean collision physics, one expects the runaway 
to happen provided that:
\begin{enumerate}
\item Core collapse, driven by mass segregation due to 2-body relaxation 
(as studied in GFR04), occurs within $\tstar\simeq 3$\,Myr {\em or}
the cluster is initially sufficiently collisional.
\item The collisional cross section is a steeply increasing function
  of a star's mass, $S_{\rm coll} \propto M_\ast^{\eta}$ with
  $\eta>1$.
\end{enumerate}
What ``sufficiently collisional'' means is not easy to establish from
simple principles. One may think that a sufficient condition is that
the collision time at the centre, for a star near the top of the IMF
(i.e., $120\,\Msun$ for our standard IMF),
$\left.\tcoll\right|_{M_{\rm max}}$ (see Eq.~\ref{eq:tcoll_ini}), be
shorter than the MS life-time for such a star. But this does not
account for the structural evolution of the cluster due to relaxation
and collisions themselves. Another condition is suggested by the
coagulation equation (\citealt{LeeMH93,LeeMH00,MG02}; see also the
simplified mathematical analysis encompassed in equations 6 and 7 of
Paper~I). In the case of strong gravitational focusing, it reduces to
$\Rstar \propto \Mstar^\beta$ with $\beta>0$ ($\Rstar$ is the stellar
radius), a condition always obeyed by MS stars (but not during the
pre-MS stage). If gravitational focusing is strong (i.e., for velocity
dispersion $\sigmaOneD > 500-1000\,\kms$), one needs $\beta>0.5$,
which is (although marginally) not satisfied by our standard VMS
$M$--$R$ relation ($\beta=0.47$, see Paper~I). In any case, the
relevance of the second condition is not clear because it stems from a
type of analysis which disregards the stellar dynamics, most
noticeably the role of mass segregation.

Fig.~\ref{fig:runaway_plane} represents the ($M_{\rm cl}$, ${\Rnb}$)
plane in which are plotted, for $W_0=3$ and $W_0=8$, the conditions
$\tccrlx = 0.12\,\trc(0) = \tstar = 3\,$Myr and
$\left.\tcoll\right|_{M_{\rm max}} = 3\,$Myr. $\tccrlx$ denotes the
relaxational core-collapse time, when collisions are absent. The value
$\tccrlx = 0.12\,\trc(0)$ is the average value yielded by {\MESSY}
when collisions are absent (see Paper~I and discussion below). We also
indicate the initial one-dimensional velocity dispersion at the
centre,
\begin{equation}
\begin{split}
\sigma_0\equiv {\sigmaOneD}_{,\rm c}(0) & \simeq 0.55\sqrt{\frac{G M_{\rm cl}}{{\Rnb}}} \\
 & \simeq 36\,\kms\, \left(\frac{M_{\rm cl}}{10^6\,\Msun}\right)^{1/2} 
\left(\frac{{\Rnb}}{1\,{\rm pc}}\right)^{-1/2}.
\end{split}
\end{equation}
This is approximately correct for any $W_0$ value. By comparing
$\sigma_0$ with the escape velocity from a stellar surface,
$500-1000\,\kms$, one can estimate whether gravitational
focusing (included in all simulations) and collisional mass loss (not
accounted for in the sticky-sphere approximation) may play an important role. 

From this diagram, it is clear that only relatively small clusters may
have $\tccrlx<3\,$Myr without being initially collisional
($\left.\tcoll\right|_{M_{\rm max}} > 3\,$Myr); for this one must have
${\Nstar}<10^6$ for $W_0=3$ and ${\Nstar}<3 \times 10^6$ for $W_0=8$.
This fact was not included in GFR04, who considered only the $\tccrlx$
condition. Consequently, there are in principle more clusters which
may form VMSs through collisions. Even though
initial conditions with collision time shorter than a few million
years are questionable, they are an interesting idealisation to pave
the way to more realistic simulations of the collisional {\em
formation} of dense clusters.

We show in Fig.~\ref{fig:runaway_plane} the initial conditions for all
simulations carried out with the standard IMF and collisions treated
as pure mergers or with the SPH-generated prescriptions. We also
indicate the outcome of each simulation, i.e., whether a VMS grew or
whether the core collapse was terminated by stellar evolution of the
massive MS stars before this could happen. We set the lower
limit for successful VMS growth at $M_{\rm VMS}\ge 400\,\Msun$. This
is a relatively arbitrary value which matters only in the case of the
four simulations with $W_0=3$, $\Nstar=10^5$ and $\Rnb=0.5\,$pc. These
runs were computed with the same code and parameters but using
different realisations of the cluster and random sequences. In three
runs, a star grew to $400-500\,\Msun$ before it left the MS; in the
fourth, only $\sim 300\,\Msun$ was reached. To determine the time at
which the runway started, $\tra$, we look for an increase of mass of a
factor 3 or more in a star more massive than $0.9\,M_{\rm max}$ (the
maximum mass in the IMF) within the last tenth of the elapsed
time. This definition may seem contrived but is required to capture
the onset of runaway as it would be naturally identified by
inspection of collision history diagrams such as
Figs~\ref{fig:coll_hist_LowColl} or
\ref{fig:coll_hist_HiColl}. In practice $\tra$ is very close to the
core collapse time one would identify by looking at the evolution of
the Lagrange radii.

We followed the merger sequence to $1000\,\Msun$ at least in most
cases but, except for a few exceptions, did not try to carry on the
simulation until the growth was terminated by evolution off the MS.
With no or little collisional mass loss, each merger brings new
hydrogen to the VMS and allows it to survive until the next merger,
see Figs~\ref{fig:2tscales300k} and \ref{fig:2tscales3M}. It is likely
that, in real cluster, the growth will saturate through some process
not included in {\MESSY}. One possibility is the depletion of the
``loss cone'', i.e., the disappearance of the stars populating the
orbits that intersect the small central volume in which the VMS move
around, see Section~\ref{subsec:loss_cone}. Another limiting mechanism
could be that the VMS cannot radiate the energy dumped into it by
collisions fast enough to keep its relatively small MS size, and
instead swells and becomes ``transparent'' to impactors, as suggested,
for other reasons, by
\citet{Colgate67}. On the MS, a VMS is not only dense enough 
to stop impactors but also resilient to disruption despite its
radiation-dominated interior. Indeed, the binding energy per unit mass
actually increases from $\sim 3$ to $\sim 4.5\,G\Msun/\Rsun$ for
masses ranging from 100 to $10^4\,\Msun$
\citep{BAC84}. In any case, we are here mostly concerned with
detecting the onset of the runaway growth and leave the difficult
prediction of its final mass for future
studies.

Our work is limited to clusters with a relatively large number of
particles: ${\Nstar}\ge 10^5$, $3\times 10^5$, for $W_0=3$, $8$,
respectively, because the MC code can only yield robust results when
there are always at least a few dozens particles in the cluster core;
otherwise, one cannot make a robust estimate of the central density,
which is required to simulate relaxation and compute the collision
probabilities.

An examination of Fig.~\ref{fig:runaway_plane} shows that, in
general, the simple expectation derived from the results of GFR04 that
VMS formation will occur when and only when the central relaxation
time is shorter than 20\,Myr is borne out by the present
simulations. However, for relatively small clusters (${\Nstar}<
3\times 10^{5}$ for $W_0=3$; ${\Nstar}< 10^{6}$ for $W_0=8$), a
shorter central relaxation time appears to be required. This is
possibly a consequence of the dearth of massive stars in the central
regions: with our standard IMF, the number of stars with
$\Mstar>50\,\Msun$ in the core of a ${\Nstar}=3\times 10^5$, $W_0=3$
or a ${\Nstar}=10^6$, $W_0=8$ cluster is $\sim 30$ or $\sim 20$,
respectively. Although, over the time required for core collapse $\sim
100\,\Msun$ stars may come to the centre from distances much larger than
the core radius, the dynamics of core collapse is driven by a small
number of particles and the applicability of the standard treatment of
relaxation (and of the notion of 2-body relaxation itself) becomes
questionable.

For large clusters, in contrast, we see that runaway VMS formation may
happen at sizes for which $\trc(0)>20\,$Myr because collisions occur
from the beginning and accelerate
core-collapse. Fig.~\ref{fig:trunaway} is a graphical attempt at
quantifying this effect. On this diagram, we plot, for all simulations
with our standard IMF which lead to VMS formation, $\tra$, the time of
runaway occurrence. Using equation~[8-123] of
\citet{BT87}, we define the initial central collision time for a star
of $120\,\Msun$ as

\begin{equation}
\tcollc(0) \equiv \left.\tcoll\right|_{M_1} = \frac{A
\left(\frac{{\Nstar}}{10^6}\right)^{-3/2} \left(\frac{R_{\rm h}}{1\,{\rm pc}}\right)^{-7/2} }{
1+B\left(\frac{R_{\rm h}}{1\,{\rm pc}}\right) \left(\frac{{\Nstar}}{10^6}\right)^{-1} }
\label{eq:tcoll_ini}
\end{equation}
where
\begin{equation}
\begin{split}
A &= \frac{4.13\times 10^{12}\,{\rm yr}}{C_n C_\sigma} \left(\frac{\langle\Mstar\rangle}{\Msun}\right)^{-1/2} \left(\frac{R_1+R_2}{\Rsun}\right)^{-2},\\
B &= \frac{22.2}{C_\sigma^2} \left(\frac{R_1+R_2}{\Rsun}\right)^{-1} \left(\frac{M_1+M_2}{\langle\Mstar\rangle}\right),
\end{split}
\nonumber
\end{equation}

\begin{equation}
\begin{split}
C_n = \frac{n(0)R_{\rm h}^3}{{\Nstar}} &= \begin{cases}
 0.385 & \text{if $W_0=3$},\\
 9.10 & \text{if $W_0=8$},
\end{cases}\\
C_\sigma = \sigmaOneD\sqrt{\frac{R_{\rm h}}{G\Mcl}} &= \begin{cases}
 0.474 & \text{if $W_0=3$},\\
 0.495 & \text{if $W_0=8$}.
\end{cases}
\end{split}
\nonumber
\end{equation}

We take $M_1=120\,\Msun$, $M_2=\langle\Mstar\rangle\simeq 0.69\,\Msun$
and $R_{1,2}=R_\ast(M_{1,2})$ according to our MS $M$--$R$
relation. When collisions are treated as perfect mergers
(``sticky-sphere'' approximation), a reasonable fit to our results for
the runaway time is
\begin{equation}
{\tra}^{-1}=(0.12\,\trc(0))^{-1}+(250\,\tcollc(0))^{-1}.
\label{eq:trunaway}
\end{equation}
Hence ${\tra}$ is approximately $0.12\,\trc(0)$, close to the
core-collapse time found in GFR04, $0.15\,\trc(0)$, for all clusters
with a relatively long initial central collision time. As $\tcollc(0)$
decreases, so does ${\tra}$, first mildly and then steeply when
$\tcollc(0)< 10^{-3} \trc(0)$. For very short $\tcollc(0)$, one has
${\tra} \simeq 250\,\tcollc(0)$. In this regime, the core collapse
appears to be driven by collisions rather than relaxation. This is
made clear by comparing the results of two simulations of a cluster
with $W_0=3$, ${\Nstar}=10^8$ (${\Npart}=3\times 10^5$) and $R_{\rm
NB}=0.05$\,pc. The first run (\Sim{K3-52}) was realised with the usual physics,
including 2-body relaxation and collisions but, for the second,
relaxation was switched off. The evolution of the Lagrange radii is
plotted for both runs in Fig.~\ref{fig:LagrRadHiColl}. With
relaxation, core collapse occurs at $\tcc\simeq 7.4\times
10^{-4}\,T_{\rm FP}$. When only collisions drive the evolution (run \Sim{K3-53}), it is
slightly slower, with $\tcc\simeq 8.9\times 10^{-4}\,T_{\rm FP}$, but
is otherwise similar.

It is somewhat surprising that the transition to collision-dominated
core collapse occurs at such a short value of $\tcollc(0)$ and that
${\tra}/\tcollc(0)$ is so long in this regime. To some extent,
this is due to our definition of $\tcollc(0)$, based on the most
massive stars in the IMF ($M_1=120\,\Msun$), a choice which may not
accurately capture the nature of the cluster evolution in the
collisional regime. Indeed stars much less massive also experience
collisions from the beginning. Using $M_1=1\,\Msun$, $R_1=1\,\Msun$ to
define $\tcollc(0)$, one finds ${\tra} \simeq 2\,\tcollc(0)$ when
$\tcollc(0)\ll \trc(0)$. In any case this relation for ${\tra}$
only holds for sticky-sphere collisions. When the SPH-inspired
prescription is used (with mass loss and merger requiring $d_{\rm
min}<R_1+R_2$) and for high velocity dispersions, collisions are less
effective at driving core collapse and ${\tra}$ is consequently
longer for a given, small $\tcollc(0)$ value, as is apparent on
Fig.~\ref{fig:trunaway}.

Although it is clearly only approximate, it is tempting to use
equation~(\ref{eq:trunaway}) to predict which conditions will lead to
collisional runaway by assuming this happens if, and only if $t_{\rm
ra}< 3$\,Myr. Therefore we have also plotted in
Fig.~\ref{fig:runaway_plane} lines indicating this
``corrected'' runaway condition. One sees that, when the velocity
dispersion is sufficiently smaller than $\Vstar$, this condition
reduces to the condition on the central relaxation time
alone\footnote{This had to be expected because the ratio of relaxation
time to collision time is $\trlx/\tcoll \sim
(\ln\Lambda)^{-1}(1+\theta)\theta^{-2}$ with
$\theta\simeq(\Vstar/\sigmaOneD)^2$.} while at high velocities, one
predicts runaway for larger cluster sizes than suggested by
$\trc$. However, the simulation results show the runaway domain for
very massive clusters to reach still slightly larger ${\Rnb}$
values. Some of this may be due to dispersion in the ${\tra}$ results
(see below) but, by inspecting the run \Sim{K3-55}, another effect is
discovered. In this case, the VMS star survives to at least
$t=4$\,Myr, despite its MS lifetime being shorter than 3\,Myr. This is
due to the combination of (1) the prescription used to set the
effective age of merger products by following their central He content
(``minimal rejuvenation'', see Paper~I), (2) the relation we use for
the value of the core He at the end of the MS as a function of the
stellar mass (Fig.~3b of Paper~I) and, (3) for this particular
simulation, the assumption of perfect merger. Because the MS lifetime
is nearly independent of the mass for $\Mstar\ge 100\,\Msun$ but more
massive stars transform a larger fraction of their mass into He on the
MS, each merger amounts to an effective rejuvenation if no mass-loss
is allowed.  Clearly collisional mass loss may change the picture in
this very special regime but we think the most important shortcoming
of the approach is to assume the VMS to be on the MS. As we will see
below, in most cases, the average time between collisions is much
shorter than the thermal timescale of a MS VMS so that it structure is
likely strongly affected by collisions and our SPH mass-loss
prescriptions would not be appropriate anyway.

The extension of the runaway domain to larger systems opens the
possibility that proto-galactic nuclei without MBH may be subject to
runaway formation of a central massive object, despite relatively long
relaxation times. Present-day nuclei have a size of a few pc. However,
using equation~(\ref{eq:trunaway}) as a guide, a compact
${\Rnb}=1$\,pc nucleus with $W_0=8$ needs to be at least as massive as
$\sim 3\times 10^9\,\Msun$ to experience collision-driven collapse in
less than 3\,Myr. For $\Rnb=3$\,pc, the minimum mass is $\sim 3\times
10^{10}\,\Msun$! So it seems that this process may operate in galactic
nuclei only if they are born very compact or concentrated. Indeed for
$W_0=10$, one would expect all clusters more compact than $\sim
2.5\,$pc to complete core collapse within 3\,Myr, independent of their
mass.

Our results show a relatively large dispersion in the runaway time
when evolution is dominated by relaxation, in which case ${\tra}$
should be very close to the core-collapse time found in GFR04,
i.e., $\tccrlx\simeq 0.15\,\trc(0)$ (with little variation from one
simulation to another). The present results indicate a slightly faster
evolution, $\tccrlx\simeq 0.12\,\trc(0)$, which is very likely the
consequence of the use of a different code\footnote{See
\citet{FB01a} for comparisons of the $\tccrlx$ values found with
different codes, in the case of the collapse of a single-mass
model. Models with a broad mass spectrum are more affected by
particle noise than single-mass clusters (at a given ${\Npart}$) and
display more intrinsic and code-to-code variation.}. The larger
intrinsic dispersion is also a property of the present code, linked to
the use of local time steps; they impose a scheme in which the
selection of the next particle pair to evolve is a random
process.

Following this overview of our results for the complete set of
standard simulations, we proceed, in the following 
subsections to a more detailed
description of the runaway process based on a few illustrative
cases. Presentation of non-standard cases, for which we varied the
IMF, $M$--$R$ relation, collision prescription, or the treatment of the
VMS, will be done in Section~\ref{subsec:nonstandard}.

\subsection{Cluster structure evolution}

\subsubsection{Missing the collisional runaway}

\begin{figure}
  \resizebox{\hsize}{!}{%
	\includegraphics[bb=22 156 585 701,clip]{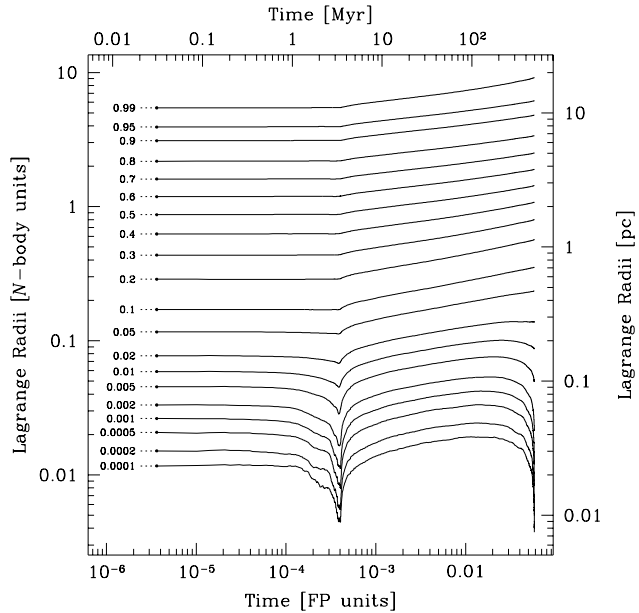} }
  \caption{%
  Evolution of Lagrange radii for model \Sim{K8-27}. For
  this cluster, the core collapse is slightly too long to allow the
  runaway process to set in. The cluster undergoes two successive core
  collapses. The first one driven by the massive MS stars, the second
  by the stellar BHs (with a mass of $7\,\Msun$) resulting from their
  evolution off the MS. For clarity, the curves have been smoothed by
  using a sliding average procedure with a (truncated) Gaussian
  kernel over 10 adjacent data points.} See also Fig.~\ref{fig:StellTypes_RA_009}.
  \label{fig:LagrRad_RA_009}
\end{figure}

\begin{figure}
\centerline{
  \resizebox{0.86\hsize}{!}{%
    \includegraphics[bb=37 296 315 694,clip]{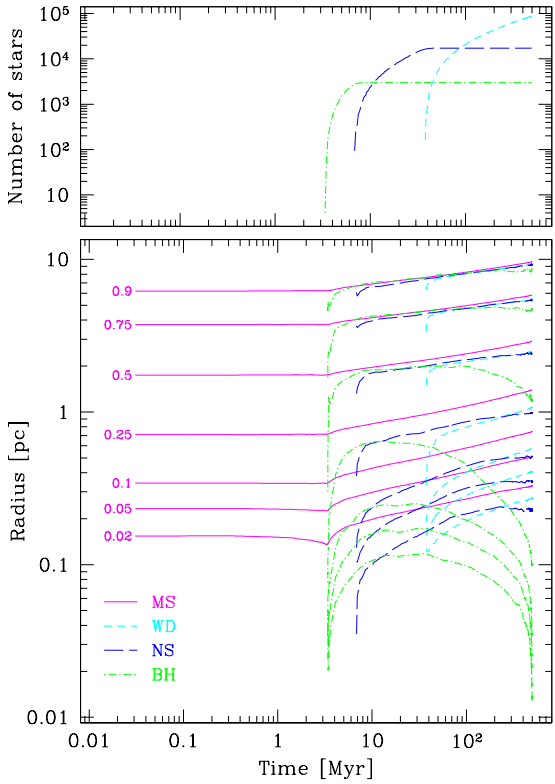}%
    }}
  \caption{%
    Evolution of the various stellar types for model \Sim{K8-27} (same
    as Fig.~\ref{fig:LagrRad_RA_009}). For this cluster, the core
    collapse is slightly too long to allow the runaway process to set
    in. The stellar types accounted for are: MS stars (MS), white
    dwarfs (WD), neutron stars (NS) and stellar BHs (BH). For
    simplicity, all WDs have $0.6\,\Msun$, all NS $1.4\,\Msun$ and all
    BHs $7\,\Msun$. Top panel: number of compact remnants as function
    of time. Bottom panel: evolution of Lagrange radii. The simulation
    was stopped as the stellar BHs underwent core collapse on their
    own.}
  \label{fig:StellTypes_RA_009}
\end{figure}

Figs~\ref{fig:LagrRad_RA_009} and \ref{fig:StellTypes_RA_009}
illustrate what occurs when the core-collapse time is longer than MS
lifetime of massive stars. For the sake of clarity, we present a case
in which the segregation-induced core collapse would have taken just
slightly longer than $\tstar=3\,$Myr. The evolution of the Lagrange
radii, plotted in Fig.~\ref{fig:LagrRad_RA_009}, indeed indicates a
first core contraction which stopped and reversed at $\tstar$. A
second collapse occurs much later, at $t\simeq 500$\,Myr. What happens
is made clear by Fig.~\ref{fig:StellTypes_RA_009}. On the top panel,
we have plotted the evolution of the number of stellar remnants; the
bottom panel is the evolution of the Lagrange radii for each stellar
type (MS: main-sequence stars, WD: white dwarfs [$0.6\,\Msun$], NS:
neutron stars [$1.4\,\Msun$], BH: stellar black holes
[$7\,\Msun$]). The first collapse stops when the massive stars turn
into BHs, causing strong mass-loss from the central region which
re-expands as its binding energy decreases. The BHs are born strongly
segregated because their massive progenitors had concentrated at the
centre. The BH distribution first re-expands but eventually
re-collapses as a result of mass segregation. From $t\simeq 40\,$Myr
onwards, they are indeed the most massive objects in the cluster. The
subsequent evolution of this population of BHs will be driven by their
binaries, either present before the second collapse, or dynamically
formed through 3-body interactions during it. The MC code used here
cannot treat binaries so we leave the study of the evolution of the
central dense BH sub-cluster out of the present work
\citep[see][]{OLFIR05}.

\subsubsection{Core collapse with runaway}

\begin{figure}
  \resizebox{\hsize}{!}{%
    \includegraphics[bb=22 156 585 701,clip]{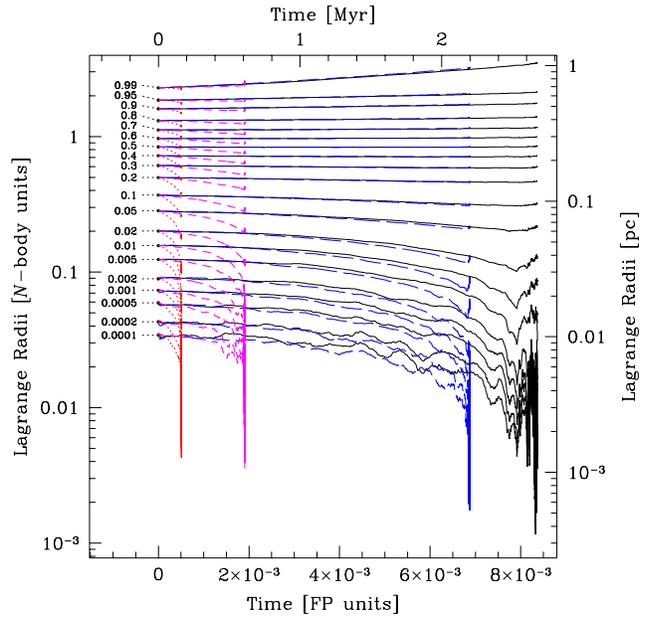}%
    }
  \caption{ Evolution of Lagrange radii for various clusters with
  $W_0=3$. From the least to the most collisional (initially): with
  solid lines (black in colour version), \Sim{K3-33}; with long dashes
  (blue) \Sim{K3-37}; with short dashes (magenta) \Sim{K3-64}; with
  dotted lines (red) \Sim{K3-61}. The top horizontal and right
  vertical axis indicate time and radius in physical units for model
  \Sim{K3-33}. The evolution of this cluster is driven by 2-body
  relaxation until late into core-collapse, at which point the
  collisions kick in. The core collapse time is thus very similar to
  the value one would obtain for point-mass dynamics.}
  \label{fig:LagrRadVarious}
\end{figure}

\begin{figure}
  \resizebox{\hsize}{!}{%
    \includegraphics[bb=22 156 585 701,clip]{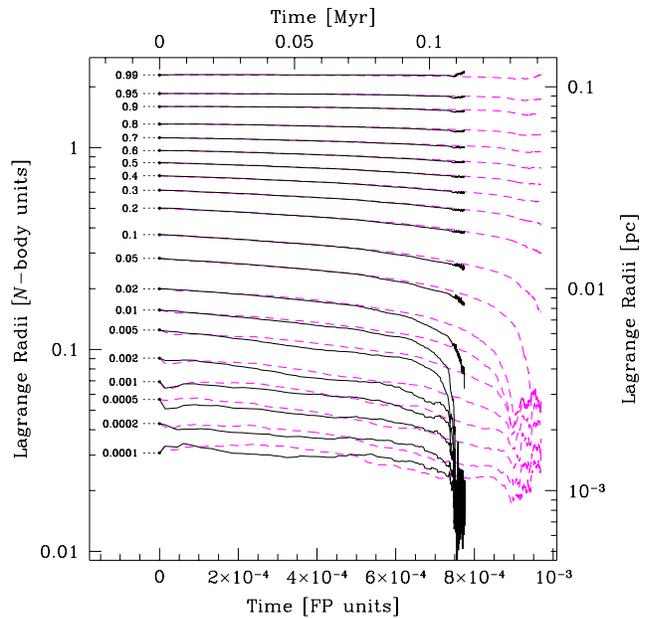}%
    }
  \caption{ Evolution of Lagrange radii for model \Sim{K3-52}. This
  cluster is strongly collisional from the beginning. The core
  collapse is driven mostly by mergers and occurs on a timescale much
  shorter (when measured in FP units). The dashed lines show the
  evolution of the same system {\em without} relaxation
  (\Sim{K3-53}). Core collapse still occurs, although on a slightly
  longer timescale.}
  \label{fig:LagrRadHiColl}
\end{figure}

\begin{figure}
  \centerline{\resizebox{\hsize}{!}{%
    \includegraphics[bb=36 151 564 690,clip]{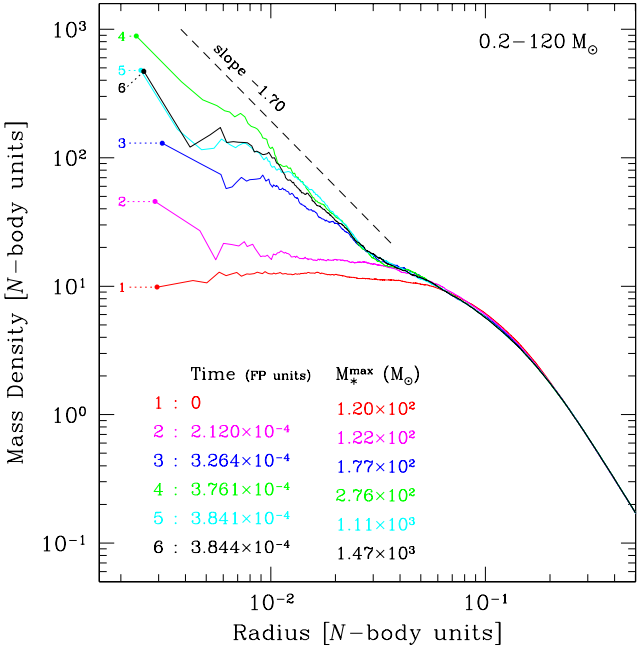}%
  }}
\caption{
  Evolution of the density profile for all stars with masses between
  $0.2$ and $120\,\Msun$ (i.e., excluding the runaway object) in
  simulation \Sim{K8-29}. For each profile in the sequence, we
  indicate the time and mass of the most massive star in the cluster
  at this time. The FP time unit is $T_{\rm FP}\simeq 4.72 \times
  10^9$\,yr. A central sub-cluster with a cuspy density profile
  forms. Its profile is compatible with a power-law of exponent
  $-1.75$ but it is not a \citet{BW76} cusp (see
  Figs~\ref{fig:prof_dens_hiM_W08}--\ref{fig:prof_disp_end_W08} and
  text).}
\label{fig:prof_dens_all_W08}
\end{figure}

\begin{figure}
  \centerline{\resizebox{\hsize}{!}{%
    \includegraphics[bb=36 151 564 690,clip]{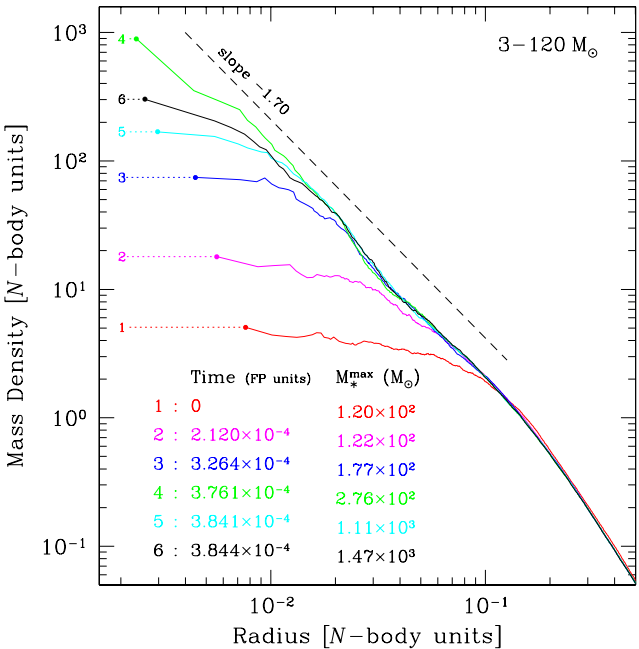}%
  }}
\caption{
  Similar to Fig.~\ref{fig:prof_dens_all_W08} but we only plot the
  density of stars with masses between $3$ and $120\,\Msun$. This
  figure confirms that the most massive stars are responsible for the
  central detached cusp seen in
  Fig.~\ref{fig:prof_dens_all_W08}. However, it seems that the
  innermost part of the density profile is more concentrated
  than a $\rho \propto R^{-1.75}$ cusp (see
  Fig.~\ref{fig:prof_dens_end_W08}).}
\label{fig:prof_dens_hiM_W08}
\end{figure}

\begin{figure}
  \centerline{\resizebox{\hsize}{!}{%
    \includegraphics[bb=36 151 564 690,clip]{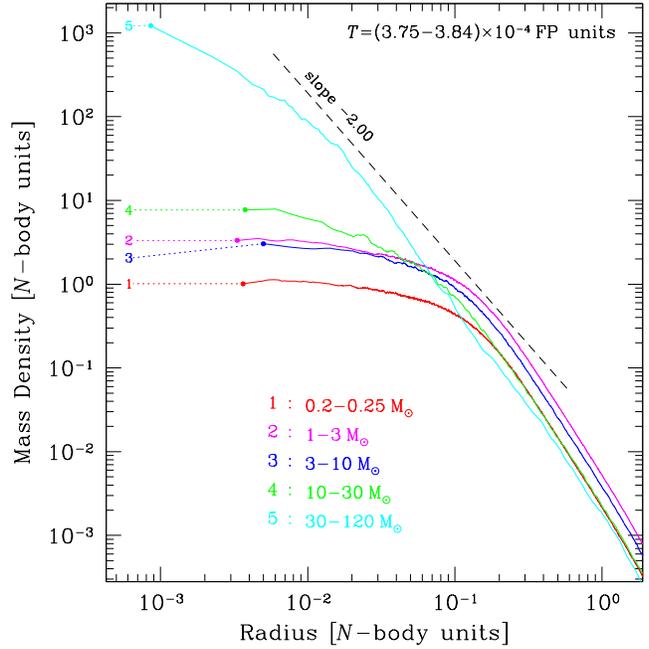}%
  }}
\caption{
 Density profiles for various stellar masses in the same simulation as
 in Figs~\ref{fig:prof_dens_all_W08} and
 \ref{fig:prof_dens_hiM_W08}. To investigate the structure of the
 cluster during the runaway phase, we have averaged the density
 profiles for 18 snapshots within the given time interval. The various
 curves correspond to different ranges in stellar mass. It is clear
 that only stars more massive than $10\,\Msun$ (with a number fraction
 of only $\sim 5\times 10^{-3}$) contribute to the formation of the
 cusp and that its profile is steeper than $\rho \propto
 R^{-1.75}$. The line with slope -2 is plotted as a guide. Note that
 the radius and density scales and ranges are different from those of
 Figs~\ref{fig:prof_dens_all_W08} and \ref{fig:prof_dens_hiM_W08}.}
\label{fig:prof_dens_end_W08}
\end{figure}

\begin{figure}
  \centerline{\resizebox{\hsize}{!}{%
    \includegraphics[bb=36 151 564 690,clip]{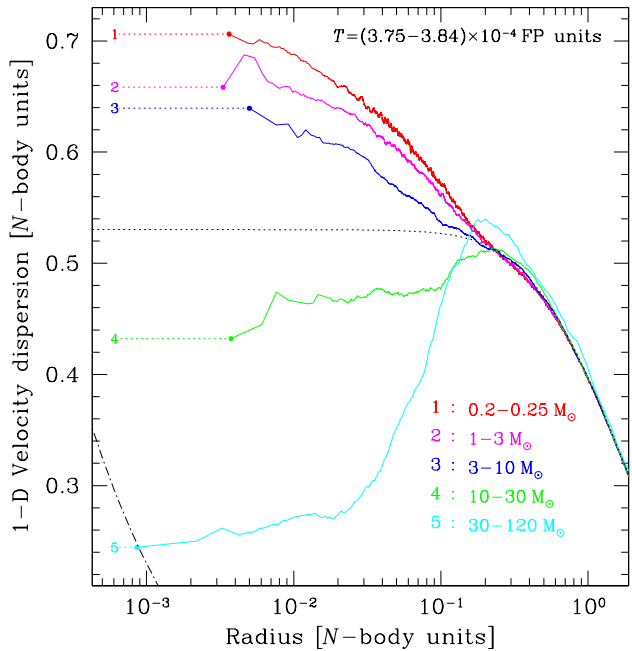}%
  }}
\caption{
  Similar to Fig.~\ref{fig:prof_dens_end_W08} but for the 1-D velocity
  dispersion $\sigmaOneD$ (plotted on a linear scale). As
  a result of (partial) energy equipartition, the most massive
  stars are considerably cooler than other stars in the central
  regions. The dotted line is the initial velocity dispersion ($W_0=8$
  King model). The dash-dotted line indicates the Keplerian dispersion
  $\sigma_{\rm K}=\sqrt{GM_{\rm VMS}/(3R)}$ corresponding to a central object with
  $M_{\rm VMS}=1000\,\Msun$. One sees that the sphere of influence of
  the runaway star (inside which $\sigmaOneD\simeq \sigma_{\rm
  K}$) is not resolved.}
\label{fig:prof_disp_end_W08}
\end{figure}

\begin{figure}
  \resizebox{\hsize}{!}{%
    \includegraphics[bb=16 151 586 693,clip]{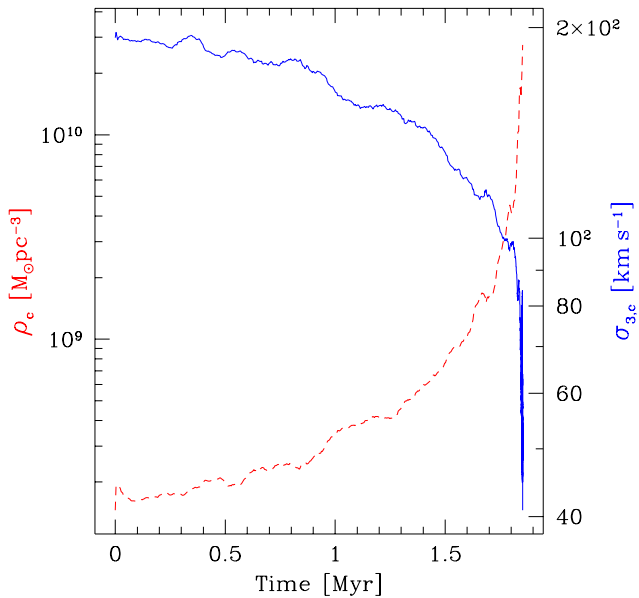}%
    }
  \caption{%
    Evolution of the central density ($\rho_{\rm c}$, dashed line, left scale)
    and 3-D velocity dispersion (${\sigmaThreeD}_{,\rm c}$, solid line, right
    scale) for model \Sim{K3-37}.  Note how the central parts
    cool down during core collapse as a result of partial
    equipartition between the massive stars and the lighter ones. }
  \label{fig:centre_RA_007}
\end{figure}

The evolution of Lagrange radii during the core collapse and runaway
VMS growth are depicted, for four $W_0=3$ cases, in
Fig.~\ref{fig:LagrRadVarious}. The first simulation is for a cluster
which not is collisional initially, i.e., from
Eq.~\ref{eq:trunaway}, with $\tcollc(0)$ ---as defined in
Eq.~\ref{eq:tcoll_ini} for $120\,\Msun$ stars--- larger than $\sim
10^{-3}\,\trc(0)$: $\tcollc(0)=3.0\times 10^6\,{\rm
yr}=0.056\,\trc(0)$. For this particular cluster, core collapse occurs in
approximately 2.5\,Myr, thus beating stellar evolution. Collisions are
very rare until the runaway starts in deep core collapse, see
Fig.~\ref{fig:coll_hist_LowColl}. Therefore, the evolution to core
collapse is very similar to what was obtained for the pure relaxation
case in Paper~I. Once the runaway has started, the
central regions re-expand as a result of the heating effect of the
accretion of deeply bound stars by the central object, as in a cluster
with a central BH destroying or ejecting stars that come too close
\citep{Shapiro77,DS82,ASFS04,BME04a,BME04b}. We cannot follow this
``post-collapse'' evolution very far because our code slows down
considerably when reaching the runaway phase, as the central relaxation
and collision times plummet. Furthermore, as will be
discussed in Section~\ref{subsec:loss_cone}, it is likely that our way of
computing collision probabilities applies correctly to the VMS only at the
beginning of its growth. 

The other simulations plotted in Fig.~\ref{fig:LagrRadVarious}
correspond to denser clusters in which collisions play an increasingly
important role in the dynamics and lead to shorter and shorter
collapse time (when measured in relaxation times). The central
collision times $\tcollc(0)$ for a $120\,\Msun$ star are $6.2\times
10^{5}\,{\rm yr}=0.014\,\trc(0)$, $8.8\times 10^3\,{\rm yr}=1.3\times
10^{-4}\,\trc(0)$ and $500\,{\rm yr}=2.9\times 10^{-5}\,\trc(0)$. As
explained in
Section~\ref{subsec:standard_overview}, in extremely dense clusters, 
core collapse is driven by collisions themselves, with relaxation
playing only a minor role. This is illustrated in
Fig.~\ref{fig:LagrRadHiColl} for another $W_0=3$ cluster. For these
runs, we have assumed that all collisions result in complete mergers. In
contrast to gravitational 2-body encounters, such mergers do not
redistribute energy between stars; they continuously dissipate orbital
energy, leading to shrinkage of all inner Lagrange radii with no
rebound.

To get a more precise description of the structure of a cluster while
it undergoes core collapse and runaway VMS formation, we look at
successive radial profiles of the stellar density and velocity
dispersion. To have a sufficient resolution of the central regions,
which are the most affected by the dynamical evolution, we use our
highest resolution simulation, model \Sim{K8-29}, with $9\times 10^6$
particles (the largest number that would fit into the memory of
available computers). The initial structure is a $W_0=8$ King
model. Fig.~\ref{fig:prof_dens_all_W08} shows the evolution of the
total stellar mass density. The core collapse is characterised by the
formation a small central density peak, growing inside the initial
constant-density core. In comparison with the core-collapse of
single-mass models, the evolution shown here is relatively
unremarkable and does not appear to proceed in a self-similar
manner. In particular, the contracting core does not leave behind a
$\rho \propto R^{-\gamma}$ region with $\gamma \simeq 2.23$
\citep[][and references therein]{FB01a,BHHM03}. The central density profile
in deep core collapse is better fitted with $\gamma \simeq 1.75$. It
may be tempting to interpret this value of the exponent by the
formation of a Bahcall--Wolf cusp \citep{BW76}, similarly to what
2-body relaxation creates in a stellar system dominated by a central
massive object. This would not be correct, however, as the following
figures make clear. When we look at the evolution of the distribution
of stars with masses in the range between 3 and 120\,$\Msun$ in
Fig.~\ref{fig:prof_dens_hiM_W08}, we see a much stronger evolution
than for the overall distribution because lighter stars, which form an
essentially static background, are so much more
numerous. Fig.~\ref{fig:prof_dens_end_W08} yields a detailed picture
of the distribution of stars in various mass ranges during the late
stages of the core collapse (while the VMS is quickly
growing). Clearly, only the spatial distribution of stars more massive
than 10\,$\Msun$ significantly changes and the central density peak is
mostly made of stars more massive than 30\,$\Msun$. These objects do
not form a $\gamma \simeq 1.75$ cusp but rather a distribution of
varying slope.

In Fig.~\ref{fig:prof_disp_end_W08} are plotted the velocity
dispersion profiles for the same ranges of stellar masses as in
Fig.~\ref{fig:prof_dens_end_W08}. Most importantly, this figure shows
that, inside a region roughly corresponding to the initial core
($\Rcore(0)=0.12$\,pc), relaxation has produced partial energy
equipartition with stars above $10\,\Msun$ having cooled to lower
velocities by heating up lighter stars. This is particularly dramatic
for the stars in the $30-120\,\Msun$ bin, which exhibit an inverse
temperature gradient. Around $R=0.2$ the velocity dispersion for these
massive stars has in fact increased above the initial value, because
the stars with higher orbital velocity are less affected by dynamical
friction and have stayed in place while others were sinking. In this
figure, we also plot the Keplerian velocity for a $M_{\rm
VMS}=1000\,\Msun$ central object, $\sigma_{\rm K}=\sqrt{GM_{\rm
VMS}/(3R)}$. Clearly, we cannot resolve the radius of influence (the
central region dominated by the central object's gravity) for any
range of stellar mass. Only within the radius of influence would one
expect a Bahcall--Wolf cusp to develop. The evolution of the central
density and velocity dispersion (for another simulation) is depicted
in Fig.~\ref{fig:centre_RA_007}. We see how the central density
steadily increases while (in contrast with the single-mass core
collapse) the velocity dispersion drops, a result of energy
equipartition. Both effects contribute to making collisions more and
more likely. Furthermore, the decrease in velocity dispersion ensures
that disruptive collisions will be unlikely unless the central
velocity dispersion is initially very high.

\subsection{Collisional runaway sequences}

\begin{figure}
  \resizebox{\hsize}{!}{
\includegraphics[bb=67 156 510 708,clip]{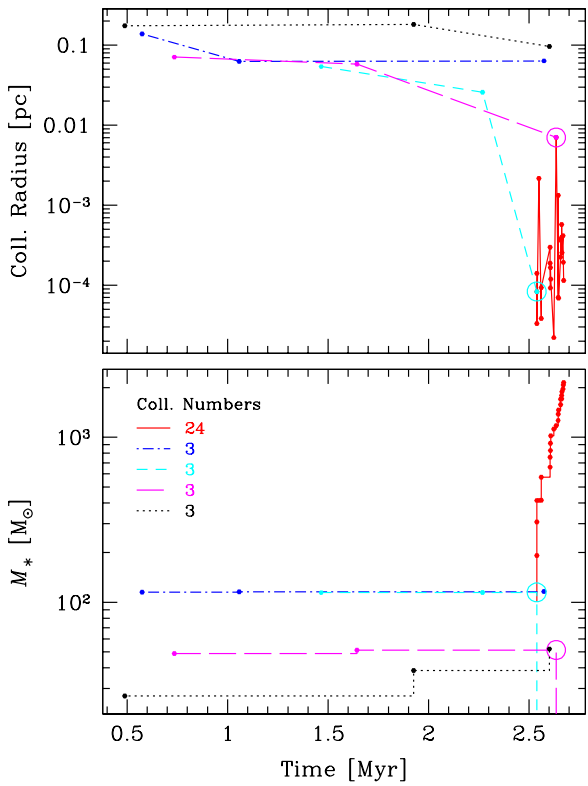} }
  \caption{ Collisional history for run \Sim{K3-33}. We represent the
  evolution of the five particles that have experienced the largest
  number of collisions. The top panel represents the radius (distance
  from the cluster centre) where the collision occurred. The bottom
  panel shows the evolution of the mass. A circled dot indicates that
  the star has merged with a larger one (usually the runaway
  star). Note that there are few collisions until $t\simeq 2.5\,$Myr,
  i.e., the moment of core collapse, at which time one star starts
  growing very quickly.}
  \label{fig:coll_hist_LowColl}
\end{figure}

\begin{figure}
  \resizebox{\hsize}{!}{%
    \includegraphics[bb=67 156 510 708,clip]{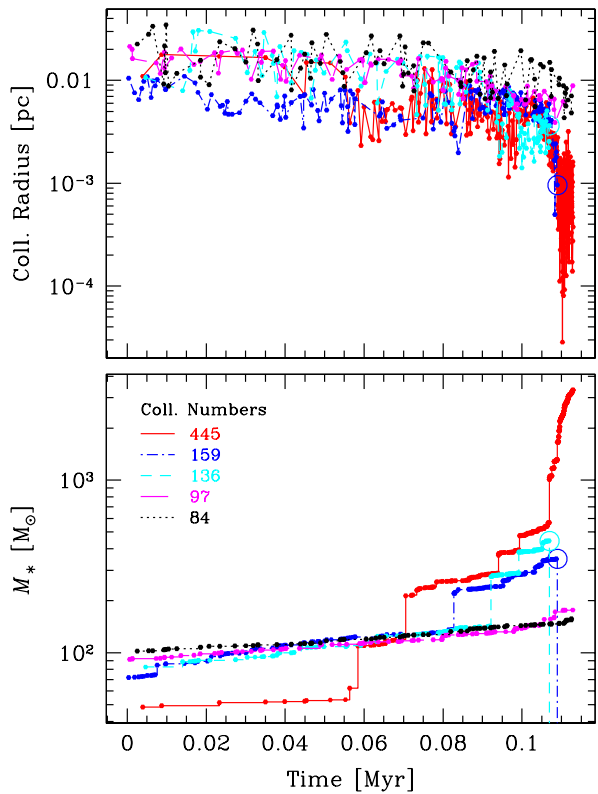}%
    }
  \caption{ Collisional history for run \Sim{K3-52}. Collisions occur
  from the beginning of the evolution and many stars grow in an
  orderly way until one detaches itself from the others and starts
  growing in a runaway fashion.  See text for explanations.}
  \label{fig:coll_hist_HiColl}
\end{figure}

We first describe results obtained in the sticky-sphere approximation
for collisions. We will then present the differences introduced by the
more realistic collision treatment based on SPH results.

A typical runaway history for a cluster which is not initially
collisional is depicted in Fig.~\ref{fig:coll_hist_LowColl}, where
we follow five stars participating in collisions, including the
runaway merger product. The parameters of the cluster are such that it
undergoes core collapse in $\sim 2.5$\,Myr. Only a few collisions
happen before deep core collapse, at which time one star detaches
itself from the rest of the population by growing in a rapid
succession of mergers. Two characteristics of this runaway growth are
its steepness and the fact that no other star experiences significant
mass increase. The initial central velocity dispersion in this cluster
is $\sigma_0 \simeq 52\,\kms$, a value for which we expect the
sticky-sphere approximation to be
accurate. Fig.~\ref{fig:coll_hist_HiColl} is an example of runaway
in a cluster which is initially strongly collisional. In this case,
many stars experience numerous collisions and start growing from the
beginning of the evolution. Nevertheless, only one of these
objects experiences a very fast growth at the moment of
(collision-driven) core collapse. Here $\sigma_0\simeq 1200\,\kms$ but
the escape velocity from a $1000\,\Msun$ VMS on the MS is
$V_{\rm esc}\simeq 3000\,\kms$, so gravitational focusing is still effective,
producing the steep increase of cross section with $M_{\rm VMS}$
required for runaway. Because relative velocities are still
significantly lower than $V_{\rm esc}$, we do not expect strong
collisional mass loss but the critical $d_{\rm min}$ for merger
becomes significantly smaller than $R_1+R_2$, making mergers less
frequent than assumed here.


\begin{figure*}
  \resizebox{11.5cm}{!}{%
          \includegraphics[bb=33 369 406 724,clip]{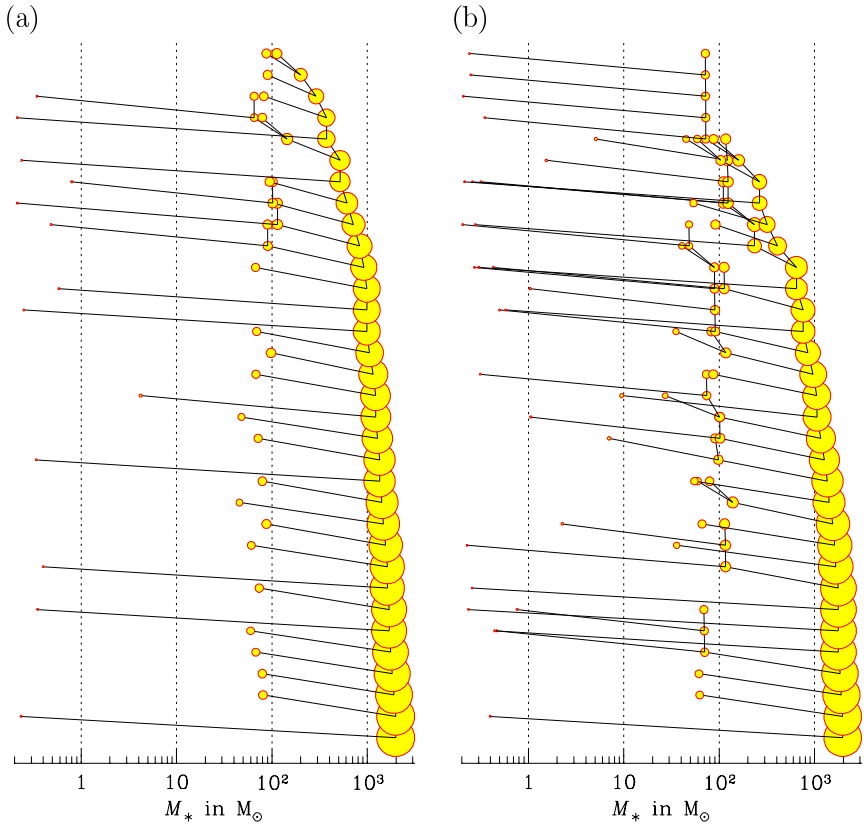}%
        }
  \hfill \parbox[b]{55mm}{%
\caption{ Merger trees for simulations of clusters with $W_0=3$,
  ${\Nstar}={\Npart}=3 \times 10^5$. Panel (a): $\Rnb=0.4$\,pc (model
  \Sim{K3-23}). Panel (b): ${\Rnb}=0.03$\,pc (model \Sim{K3-9}). We
  follow the growth of the runaway star to $\sim 2000\,\Msun$. The
  horizontal axis indicates the mass of stars taking part in the
  collisions. The sequence of mergers is represented in the vertical
  direction from top to bottom. The right ``trunk'' is the growing
  VMS. The radius of the disks is proportional to that of the
  corresponding star. In case (a), the cluster is not initially
  collisional (the collision time for a $120\,\Msun$ star is much
  larger than its MS life-time) and most collisions occurs in deep
  collapse and feature stars of mass $60-120\,\Msun$ that have
  segregated to the centre. Most of them are not themselves merger
  products. In case (b), the cluster is initially collisional and,
  although the runaway also occurs in core collapse, most stars
  contributing to it have experienced earlier collisions.}
\label{fig:2trees300k}
}
\end{figure*}

\begin{figure*}
  \resizebox{11.5cm}{!}{%
          \includegraphics[bb=33 369 406 724,clip]{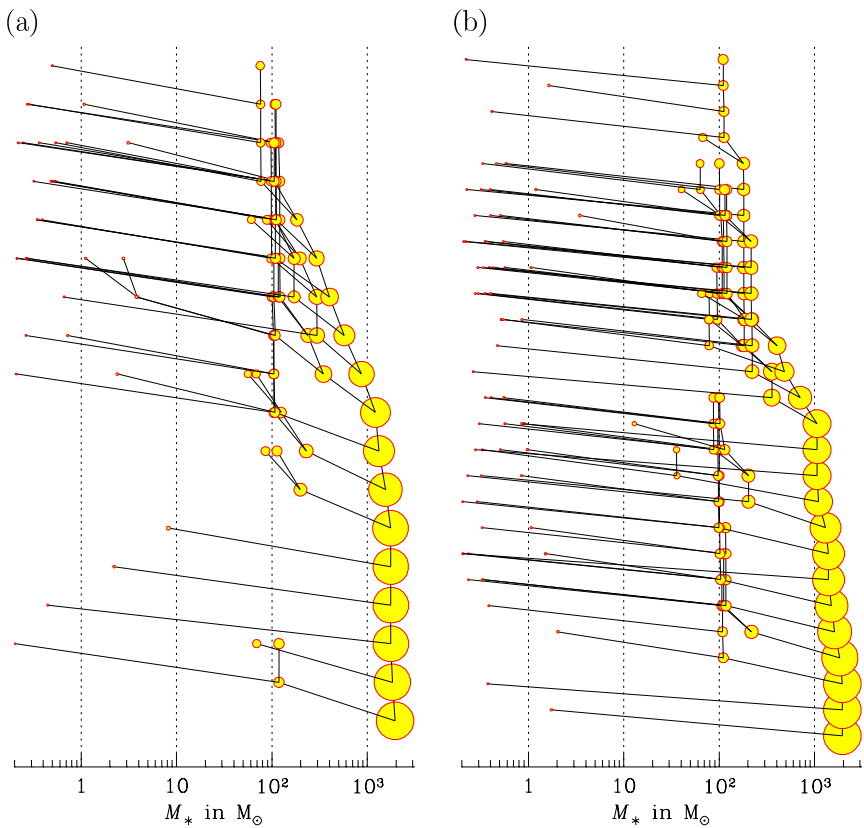}%
        }
  \hfill \parbox[b]{55mm}{%
\caption{ Merger trees for the simulation of a cluster with $W_0=3$,
  ${\Nstar}={\Npart}=3 \times 10^6$. Panel (a): $\Rnb=0.2$\,pc (model
  \Sim{K3-37}). Panel (b): ${\Rnb}=0.1$\,pc (model \Sim{K3-36}). We
  follow the growth of the runaway star to $\sim 2000\,\Msun$. Both
  clusters have initial central collision times for $120\,\Msun$ stars
  shorter than 3\,Myr, hence the relatively complex merger trees. The
  more compact cluster, case (b), is more collisional which explains a
  stronger contribution of low-mass stars.}
\label{fig:2trees3M}
}
\end{figure*}

\begin{figure*}
  \resizebox{\hsize}{!}{%
          \includegraphics[bb=35 390 495 601,clip]{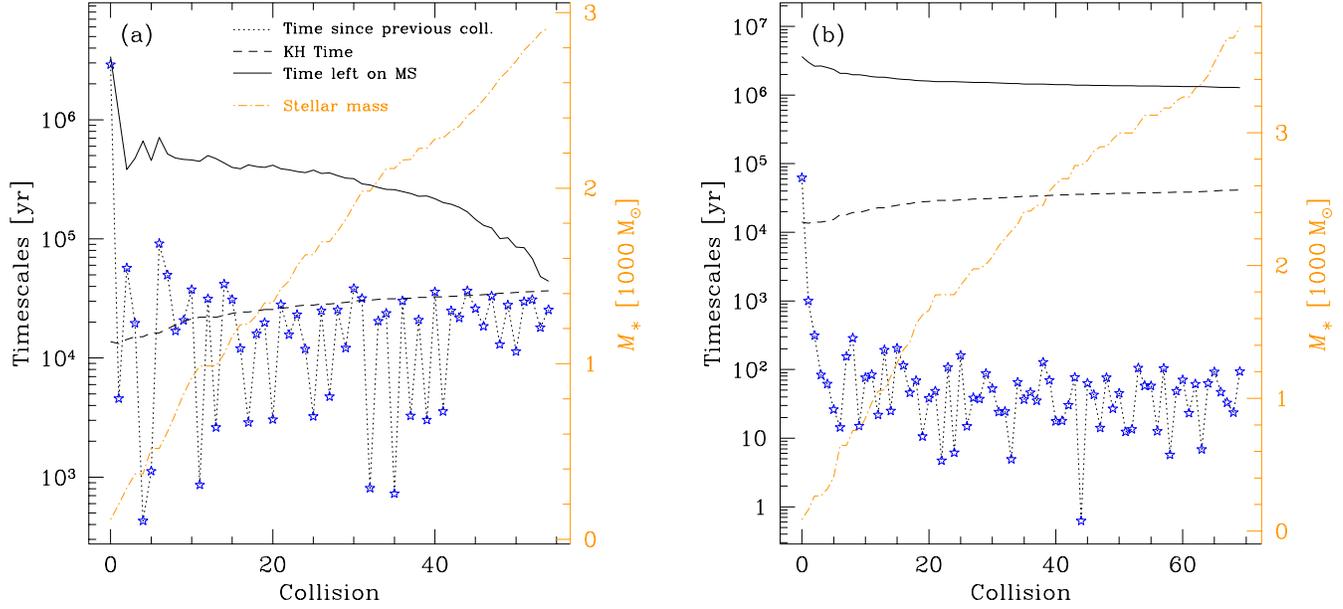}%
        }
 \caption{ Evolution of mass and various timescales during the runaway
 for the same simulations as in Fig.~\ref{fig:2trees300k}. We plot
 the time between successive collisions, an estimate of the
 Kelvin-Helmholtz timescale $\tKH$ of the collision product (assuming
 normal MS $M$--$R$ relation and luminosity), the time left until
 exhaustion of hydrogen at the centre (from our simple ``minimal
 rejuvenation'' prescription), as well as the mass of the star (right
 scale). No use is made of $\tKH$ during the simulations. We plot it
 here to show than in the vast majority of cases, the interval between
 collisions is (much) shorter than $\tKH$ so that the star would
 probably be out of thermal equilibrium, with a swollen structure.}
\label{fig:2tscales300k}
\end{figure*}

\begin{figure*}
  \resizebox{\hsize}{!}{%
          \includegraphics[bb=35 390 495 601,clip]{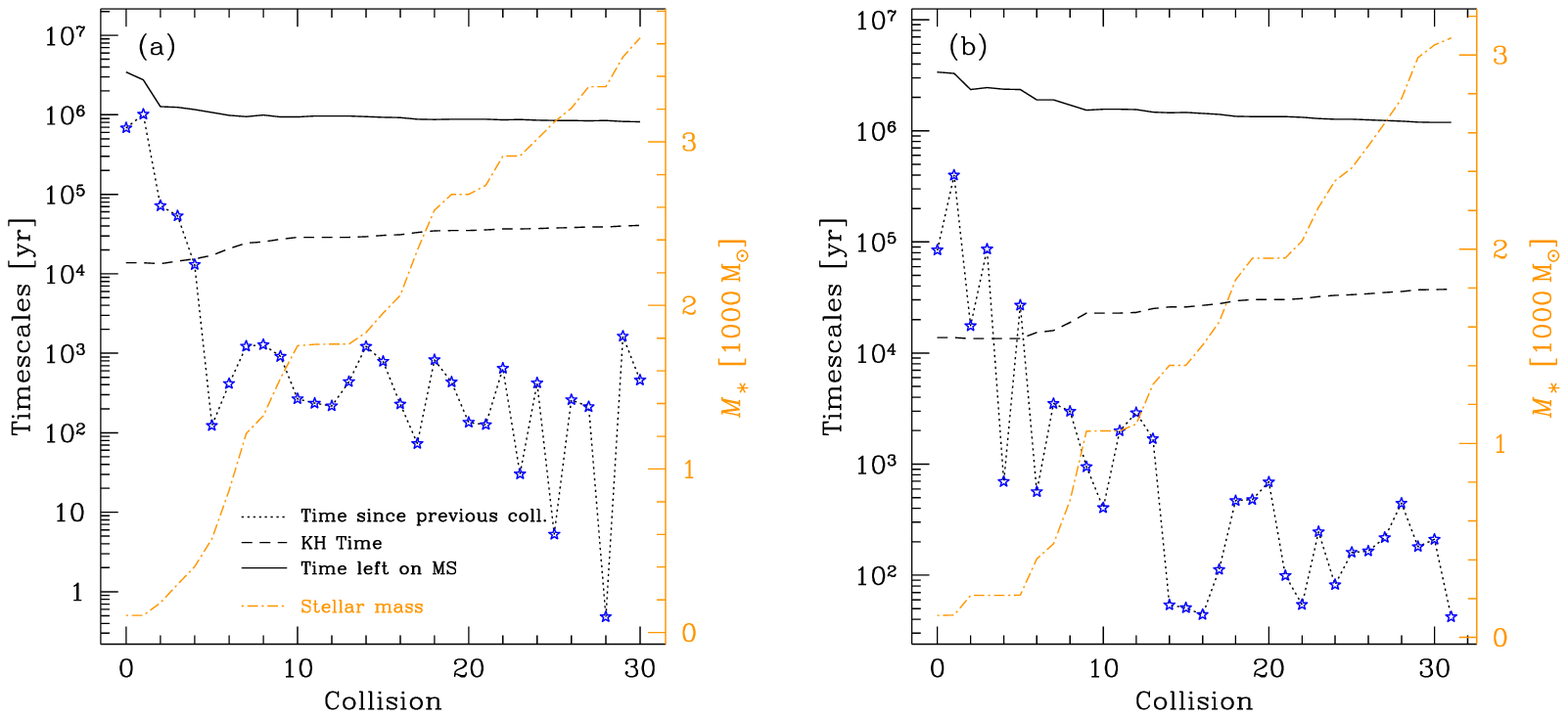}%
        }
 \caption{ 
   Mass and timescales evolution during the growth of the runaway
   star for the same simulations as in Fig.~\ref{fig:2trees3M}. See
   caption of Fig.~\ref{fig:2tscales300k} for explanations.}
\label{fig:2tscales3M}
\end{figure*}

To study which stars contribute to the build-up of the VMS, we draw
merger trees. Four examples are presented in Figs~\ref{fig:2trees300k}
and \ref{fig:2trees3M}. The first is for two clusters with $W_0=3$ and
${\Nstar}={\Npart}=3\times 10^5$, the second for $W_0=3$ and
${\Nstar}={\Npart}=3\times 10^6$. In each figure, panel (b) shows the
case of a smaller, initially more collisional cluster. In all cases,
we follow the VMS growth up to $2000\,\Msun$. The cluster in
Fig.~\ref{fig:2trees300k}a has $\Rnb=0.4$\,pc, leading to collapse in
$\tcc=2.9$\,Myr, just in time to produce collisional runaway before
the massive stars would have left the MS. In this situation, typical
of clusters which first become collisional in deep collapse, the
merger tree has a very simple structure, with most branches lacking
sub-structure: the stars that merge with the VMS (the trunk) are not
themselves collision products. Furthermore, a large majority of these
impactors have masses around the tip of the IMF, between 60 and
120\,{\Msun}. These stars add up to 92\,\% of the VMS mass. The
distribution of the number of contributing stars (the tips of the
branches, i.e., stars that are not themselves collision products) is
bimodal with one peak between $0.2$ and $1\,\Msun$ and another
covering the $40-120\,\Msun$; it reflects the MF in the central
regions (GFR04). This is to be contrasted with
Fig.~\ref{fig:2trees300k}b, obtained for a much more collisional
cluster with ${\Rnb}=0.03$\,pc, which enters the runaway phase at
$t\simeq 0.064$\,Myr. Here most stars merging with the VMS were
themselves produced by previous mergers and the contribution of
lighter stars is larger, although the $60-120\,\Msun$ range still
contributes 78\,\% of the VMS mass. The cause of the difference is
that, in the more compact cluster, collisions occur from the
beginning, when the cluster is not segregated yet; in the larger
cluster, virtually all collisions take place in the high-density core
formed by the concentration of high-mass stars. The two clusters in
Fig.~\ref{fig:2trees3M}, with $3\times 10^6$ stars and sizes of
${\Rnb}=0.2$\,pc and 0.1\,pc are both relatively collisional, as
required for this $W_0$ value and this number of stars to experience
core collapse in less than 3\,Myr (see Fig.~\ref{fig:runaway_plane});
thus, they exhibit complex merger trees. Again, in the more compact
cluster, low-mass stars play a larger but not predominant role, with
some 90\,\% of the VMS mass originating in stars more massive than
60\,{\Msun}. \Comment{MARC}{MARC}{Make sure these merger trees are
correct! The look weird...} Merger trees for $W_0=8$ clusters exhibit
the same general characteristics and variety. In particular, the same
kind of bimodal distribution in the masses of the contributing stars
can be observed, with more stars in the $0.2-1\,\Msun$ range for
clusters with shorter initial $\tcoll/\trlx$.


It is instructive to examine the various timescales involved in the
runaway process. This is done in Figs~\ref{fig:2tscales300k} and
\ref{fig:2tscales3M} for the same four simulations as in
Figs~\ref{fig:2trees300k} and \ref{fig:2trees3M}. In these
diagrams, we follow the VMS, merger after merger, making use of two
ordinate axis, one to indicate its mass, the other to monitor
important timescales. Those are the time to the next collision
$\Delta \tcoll$, the time left on the MS $\Delta t_{\rm MS}$ (if
the star were left to evolve without further collision), and the MS
thermal relaxation time, i.e., the Kelvin-Helmholtz (K-H) time. The
latter is approximated here by $\tKH=G\Mstar^2(\Rstar
L_\ast)^{-1}$ where $L_\ast$ is the luminosity of the star (close to
the Eddington limit for VMS; see, e.g.,
\citealt{BHW01,Schaerer02}). From 100 to $10^4\,\Msun$, $\tKH$ 
increases slowly from $\sim 10^4$ to $6 \times 10^4$\,yr. Here
we do not truncate at $M_{\rm VMS}=2000\,\Msun$, but show instead the full
extent of the simulations. However, in all but the first case
(Fig.~\ref{fig:2tscales300k}a), we stopped the simulation before the
VMS left the MS, when the average interval between mergers was
still much shorter than the remaining MS lifetime.

Because it corresponds to a core-collapse time just slightly shorter
than 3\,Myr, the first simulation of the four is one of the
exceptional few for which we could afford to integrate until stellar
evolution caught up with collisions, as indicated by the convergence
of the $\Delta t_{\rm coll}$ and $\Delta t_{\rm MS}$ curves. In most
other cases, $\Delta t_{\rm MS}$ still exceeded $\Delta t_{\rm MS}$ by
orders of magnitude ---suggesting many more collisions to come--- when
we decided to terminate the simulation. From the merger trees and the
curve showing $M_{\rm VMS}$ as function of the collision number, there
is only little sign of a decrease with time, as massive stars that
have segregated to the centre are progressively depleted. Such trend
can be seen for the runs with $3\times 10^5$ stars (in particular if
one draws the complete merger tree past $M_{\rm VMS}=2000\,\Msun$),
but is not perceptible for a cluster of $3\times 10^6$ stars whose central
parts constitute a larger reservoir of massive stars. Indeed among
$3\times 10^5$ stars forming a $W_0=3$ cluster, there is only about
$1600\,\Msun$ of stars more massive than $60\,\Msun$ in the core.

An essential result is that, in the vast majority of cases the K-H
time is (much) longer that the average time between successive
collisions. This means that the VMS is unlikely to relax to thermal
equilibrium, i.e., to the MS after it has merged with an impactor and
before the next collision. In principle, then, it may be kept in a
swollen state by repeated collisions and our assumptions of MS
$M$--$R$ relation and continuous nuclear burning (to set $\Delta
t_{\rm MS}$) are probably incorrect during runaway growth. This
suggests a very different picture concerning the evolution of the VMS
in the runaway phase and the termination of the growth. The VMS may
grow larger and larger until it is so diffuse that it lets the smaller
stars fly through it without the ability to stop them, hence bringing
back the ``transparency saturation''
\citep{Colgate67,LS78}. Another important effect is that the
central nuclear reactions are very likely switched off in a merger
product which expands as a result of shock heating
\citep[e.g.,][]{LWRSW02,LombardiEtAl03}, so the stellar evolution of the VMS may
be suspended. Of course these effects will persist only as long as
$\Delta \tcoll < \tKH$. When the collisions eventually become rarer
---from exhaustion of stars to collide with or because stellar
evolution from normal massive stars drives core re-expansion at
$t\simeq 3\,$Myr--- the VMS returns to the MS and to nuclear
burning. Given the complexity of interplay between repeated collisions
and VMS structure (itself determining the probability and outcome of
further collisions), it seems impossible to make any simple prediction
as to what the final VMS mass would be when it eventually leaves the
MS. In future works, this situation may be studied by combining a
stellar dynamics code with on-the-fly SPH simulation to follow the VMS
structure collision after collision \citep{BN04}. For the time being,
these aspects have to be considered as one central source of
uncertainty regarding the outcome of the runaway phase.

\begin{figure}
  \centerline{\resizebox{\hsize}{!}{%
    \includegraphics[bb=16 159 568 694,clip]{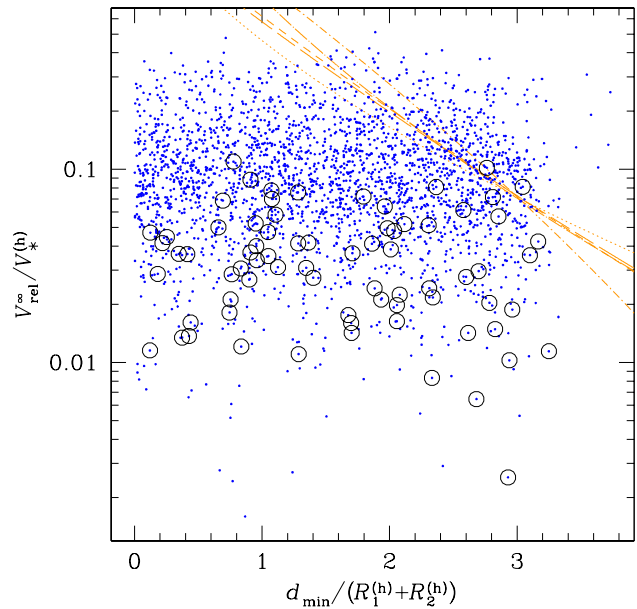}%
  }}
\caption{
  Collision parameters for a cluster with relatively low velocity
  dispersion (same simulation as in Figs~\ref{fig:2trees3M}a and
  \ref{fig:2tscales3M}a). For each collision (dot), we indicate the
  ``impact parameter'', in units of the stellar half-mass radii, and
  the relative velocity at large separation, in units of
  $\Vstar^{({\rm h})}$, see Eq.~13 of Paper~I). Circled dots indicate
  mergers involving a star with a mass larger than $120\,\Msun$ (i.e.,
  the runaway object). The lines indicate the condition for merger
  (Eq.~13 of Paper~I) for $[M_1,M_2]/\Msun= [1,1]$ (solid line),
  $[1,10]$, (dots), $[0.1,1]$, (short dashes), $[0.1,60]$, (long
  dashes) and $[60,60]$, (dot - dash). Only below these lines should
  the collisions result in mergers.}
\label{fig:coll_param_lowV}
\end{figure}

\begin{figure}
  \centerline{\resizebox{\hsize}{!}{%
    \includegraphics[bb=100 245 545 687,clip]{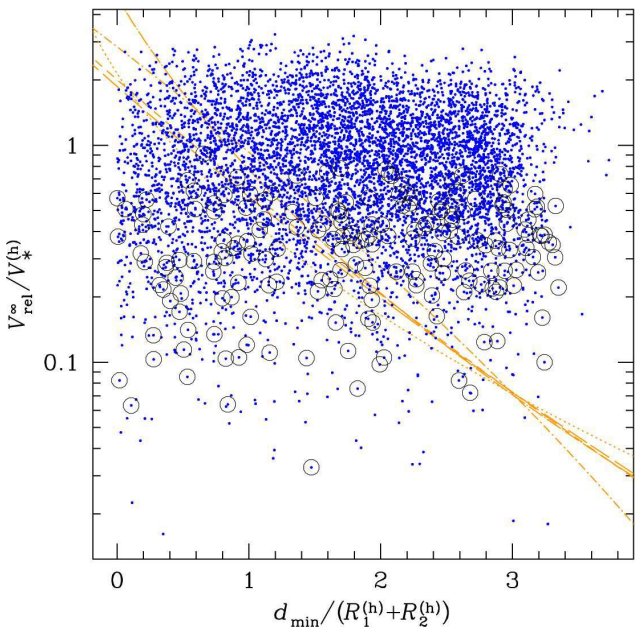}%
  }}
\caption{
  Same as Fig.~\ref{fig:coll_param_lowV} but for a cluster with high
  velocity dispersion: model \Sim{K3-55}. For this case, a large fraction of
  collisions would actually be fly-bys and the assumption of perfect
  mergers is questionable.}
\label{fig:coll_param_highV}
\end{figure}

\begin{figure*}
  \centerline{\resizebox{\hsize}{!}{%
          \includegraphics[bb=34 343 756 650,clip]{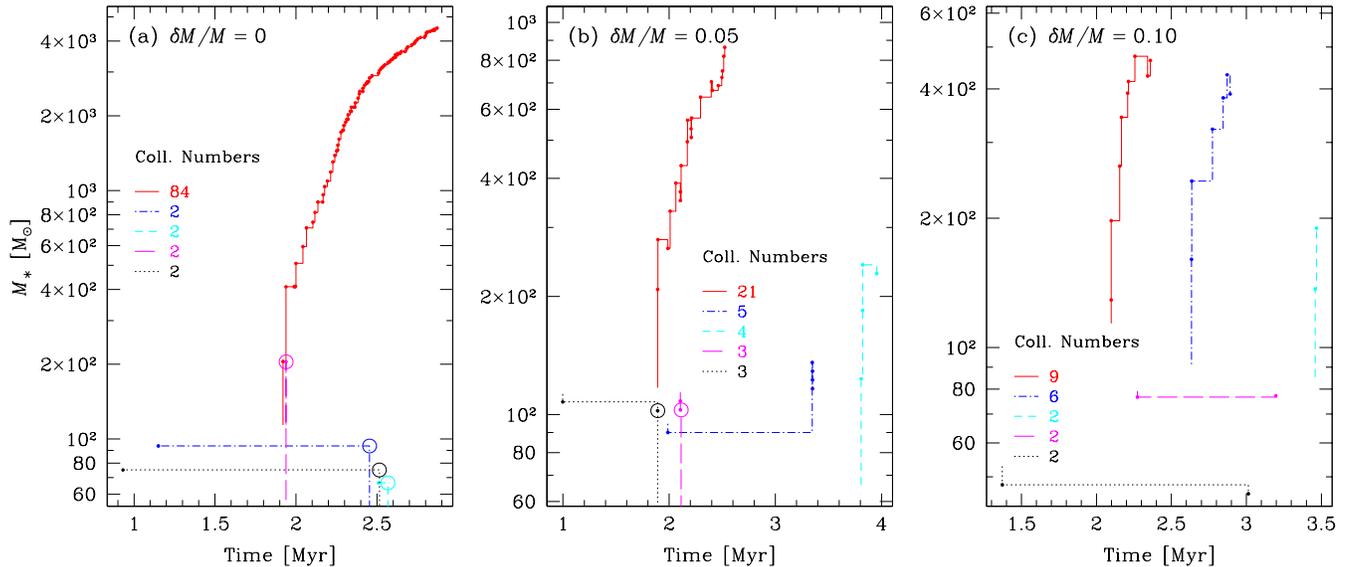}%
        }}
%
\caption{
  Role of mass-loss in the runaway process. For the three simulations
  presented here, we used the same initial model with $W_0=3$,
  ${\Nstar}=3\times 10^5$ and ${\Rnb}=0.3$\,pc (models \Sim{K3-13},
  \Sim{K3-14} and \Sim{K3-15}). All collisions are assumed to result
  in merger with a constant fractional mass loss $\delta_M\equiv
  \delta M/M$, i.e., $M_{\rm merger}=(1-\delta_M)(M_1+M_2)$ where
  $M_1$, $M_2$ are the masses of the colliding stars. This
  prescription is artificial in that a collision with a light star
  will cause more damage than one with a more massive star. For case
  (a), with no mass loss, the simulation was interrupted before the
  runaway object evolved off the MS because mergers constantly
  rejuvenate the star. With some mass loss, stellar evolution can
  catch up with collisional growth and terminate it. For simplicity,
  in these cases, we did not assume an IMBH was formed at the end of
  the VMS MS, but a $7\,\Msun$ BH instead, which was not allowed to
  collide with other stars.}
\label{fig:3massloss}
\end{figure*}

\begin{figure}
  \centerline{\resizebox{\hsize}{!}{%
    \includegraphics[bb=24 148 563 689,clip]{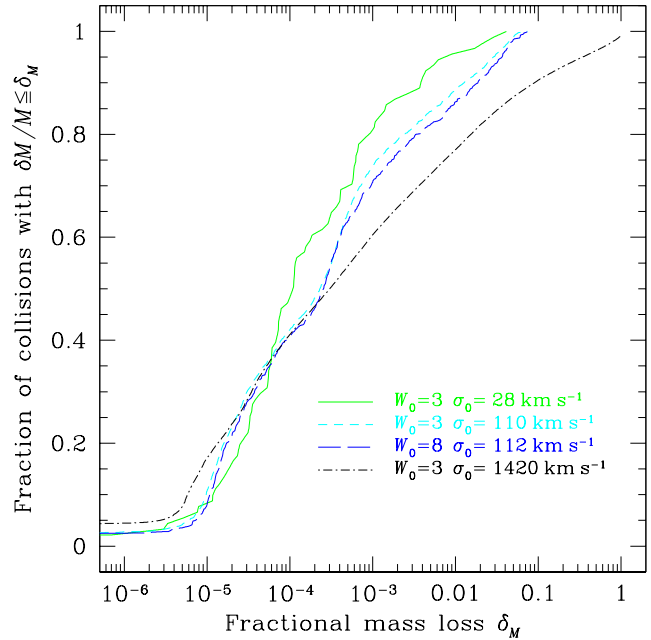}%
  }}
\caption{
Cumulative distribution of fractional mass loss in collisions for four
cluster simulations using the SPH prescription. In all four cases,
runaway VMS formation happened. The average fractional mass losses for
these runs are 0.2, 0.4, 0.5 and 4.9\,\%.}
\label{fig:dist_dMcoll}
\end{figure}

Putting aside the possible complications due to the non-MS structure
of the VMS, we can look at the changes in the runaway process
introduced by a more realistic treatment of collisions between MS
stars, as allowed by the SPH
results. Figs~\ref{fig:coll_param_lowV} and
\ref{fig:coll_param_highV} show when and why the sticky sphere
approximation may become questionable. These diagrams show the impact
parameter and relative velocity for all collisions that happened in
two cluster simulations, one with $\sigma_0\simeq 110\,\kms$
(Fig.~\ref{fig:coll_param_lowV}), the other with $\sigma_0\simeq
635\,\kms$ (Fig.~\ref{fig:coll_param_highV}). For lower velocity
dispersion, a large majority of encounters happened with parameters
leading to merger according to Eq.~13 of Paper~I. In particular, all
collisions but one in which the runaway object took part were in this
regime. They had relative velocities smaller than $0.1\,\Vstar$ and
would have produced less than $1-3$\,\% mass loss
\citep{FB05}. Consequently, treating {\em all} collisions as perfect 
mergers is certainly a very good approximation for system with
comparable or lower velocity dispersion, including any reasonable
globular cluster. In the high-velocity situation, a significant
fraction of the collisions (featuring the runaway object or not)
should have left the stars unbound rather than produced a single
merged object, suggesting that using the SPH prescription to set the
impact parameter for merger may affect the results. On the other hand,
most collisions occurred at less than $\Vstar$ and would have produced
no more than a few \% mass loss.

To assess the impact of mass loss on the runaway mechanism, we have
performed a few runs for which we assumed that all collisions result
in a merger with some constant fractional mass loss $\delta_M$, varying
the value of $\delta_M$. We show runaway growth sequences for
$\delta_M=0$, 0.05 and 0.10 in Fig.~\ref{fig:3massloss}. One sees that
$\delta_M \ge 0.10$ is required to prevent the growth of a star more
massive than a few $100\,\Msun$. As typical impactors have a mass
around $M_{\rm imp}\approx 100\,\Msun$, it is obvious why a VMS of
$1000\,\Msun$ can only be grown through a sequence of merger if
$\delta_M<0.1$. The fact that we assume all mass lost to originate
from the hydrogen envelopes also shortens the remaining MS lifetime of
the collision product and prevents reaching $M_{\rm VMS}\simeq
M_{\rm imp}/\delta_M$. 

SPH simulations show mass loss of at most a few percent for
$\Vrel<\Vstar$, so reaching $\delta_M\approx 0.1$ (on average) would
require really extreme conditions. The cluster with the highest
velocity dispersion for which we have performed a simulation using the SPH
prescriptions is model \Sim{K3-51}, with
$\sigma_0\simeq 1270\,\kms$. It has the same parameters as the
sticky-sphere run \Sim{K3-61} and we obtain the runaway VMS growth
in a qualitatively similar fashion, although about 4 times more
collisions are required to attain a given $M_{\rm VMS}$. The most
noticeable difference is that the time required to enter the runaway
phase ${\tra}$ is also about 4 times longer than in the
sticky-sphere approximation (see left-most open triangle in
Fig.~\ref{fig:trunaway}). This is obviously a consequence of the
collisions being about 4 times less efficient at driving collapse in
this collision-dominated cluster. 
In cases with lower velocity dispersion, using the SPH prescription
make very little change to the conditions required to achieve fast
collapse and collisional runaway or to the runaway itself. 

In Fig.~\ref{fig:dist_dMcoll}, we plot the cumulative fractional
mass loss for all collisions that took place in four simulations for
clusters with initial central velocity dispersions ranging from
$\sigma_0=28$ to 1420\,{\kms}. We see that all collisions yield
$\delta_M<10$\,\% for $\sigma_0 \lesssim 100\,\kms$ with an average
$\langle \delta_M\rangle<1$\,\%. The cluster with the highest
$\sigma_0$ experiences a few completely disruptive collisions but only
involving stars less massive than $2\,\Msun$. Even in this extreme
case, $\langle \delta_M\rangle$ is only of order 5\,\%. For collisions
involving a star more massive than $120\,\Msun$, the average mass loss
is as low as 0.3\,\%.

The only case for which we find a qualitatively different evolution when using
SPH-generated collision recipes instead of pure mergers, is for
$W_0=3$, $\Nstar=10^8$ and $\Rnb=0.2\,$pc. In \Sim{K3-55}, assuming
pure mergers, we found runway, thanks to collisional rejuvenation that
kept the VMS on the MS for more than $4\,$Myr. With collisional mass
loss and a large fraction of collisions not resulting in mergers, the
(orderly) growth of stars stopped at $t\simeq 3.5$\,Myr, before any
runaway star detached from the mass spectrum (runs \Sim{K3-65} and
\Sim{K3-56}).

Because a realistic treatment of collisions has so little impact on
most results pertaining to the cluster evolution and runaway VMS
growth, it is no surprise that we find them essentially unchanged if,
instead of our SPH-based prescription, we implement the simple fitting
formulae published by \citet{Rauch99}, which were derived from a
smaller set of SPH simulations.

\subsection{Non-standard simulations}
\label{subsec:nonstandard}

\subsubsection{Other IMFs}
\label{subsec:otherIMFs}

\begin{figure}
  \centerline{\resizebox{0.75\hsize}{!}{%
    \includegraphics[bb=178 175 460 691,clip]{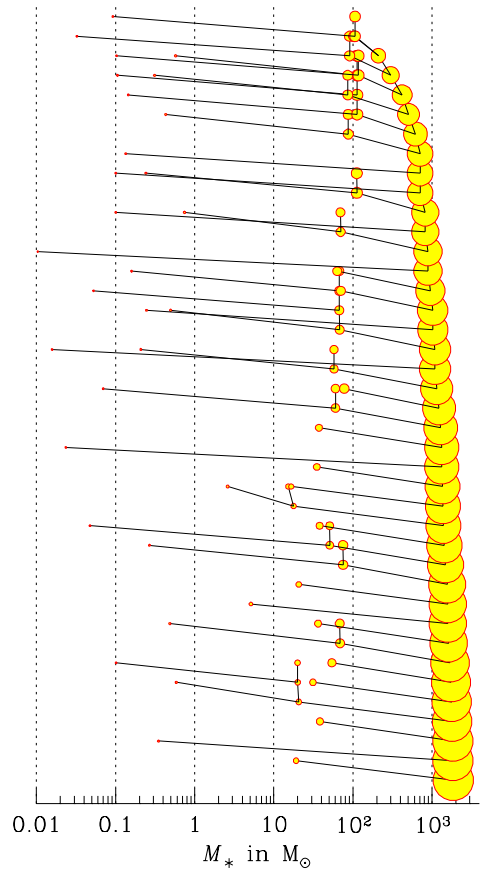}%
  }}
\caption{ 
Merger tree for the runaway growth of a VMS in model \Sim{K3-61}, a
cluster with steep Kroupa IMF extending from $0.01$ to $120\,\Msun$ (see
text).}
\label{fig:merg_tree_Kr}
\end{figure}

GFR04 established that, for any realistic IMF and all cluster
models we tried for which $\trc(0)>0$, the segregation-driven core
collapse occurs at $\tccrlx\simeq 0.15\,\trc(0)$. By ``realistic'' IMF we
mean one that is sufficiently broad, with $\mu\equiv M_{\ast,\rm
max}/\langle M_\ast\rangle>50$, and not too steep. This includes the
``Kroupa'' IMF \citep{KTG93}, which contains significantly fewer
high-mass stars than the Salpeter IMF (with $d{\Nstar}/{\rm d}M_\ast
\propto M_\ast^{-2.7}$ for $M_\ast > 1\,\Msun$). Although this is
actually an effective IMF for stars forming the Galactic field, with
clusters probably having a Salpeter-like IMF beyond $1\,\Msun$
\citep{KW03,WK04}, we used it in order to investigate the possible
effects of a smaller fraction of massive stars. We find that the
runaway growth of a VMS happens essentially in the same way as with
the Salpeter IMF. The only noticeable, and not unexpected, difference
is that because stars around $100\,\Msun$ are so much rarer, the
average impactor mass decreases much more quickly during the VMS
growth, as can be seen in the merger tree shown in
Fig.~\ref{fig:merg_tree_Kr}.

We have also simulated clusters with a reduced stellar mass range. The
single-mass case lacks realism and was only considered in Paper~I as a
test-case to compare with \citet{QS90}.

If the Salpeter IMF only extends from $0.2$ to $10\,\Msun$, the
average stellar mass is $0.58\,\Msun$ and $\mu=10/0.58\simeq 17$. The
results of GFR04 suggest $\tccrlx\simeq 0.7\,\trc(0)$ for $W_0=3$. For
such a cluster, we find ${\tra}\simeq 1.2\,\trc(0)$ (run
\Sim{K3-20}). The 10\,{\Msun} stars leave the MS after
$\tstar=25$\,Myr; these two nearly equally longer timescales combine
to give conditions for runaway in the $({\Nstar},\,{\Rnb})$ plane that
are very close to those for our standard IMF. Hence, the details of
the upper end of the IMF have little effect on the conditions for the
onset of runaway collisions. Of course, if the maximum stellar mass in
the IMF is lower, a larger number of mergers are required to grow a
VMS but there is about 3 times more mass in stars in the range
$5-10\,\Msun$ in a Salpeter IMF truncated at $10\,\Msun$ than in the
range $60-120\,\Msun$ if the IMF extends to $120\,\Msun$. In
simulation
\Sim{K3-20}, the cluster entered the runaway regime at ${\tra}\simeq
\tccrlx=20$\,Myr, allowing a VMS to reach $1500\,\Msun$ when we stopped 
the computation. 

\subsubsection{Conditions for collisional runaway in MGG-9}
\label{subsec:MGG}

\begin{figure}
  \centerline{\resizebox{\hsize}{!}{%
    \includegraphics[bb=32 158 590 696,clip]{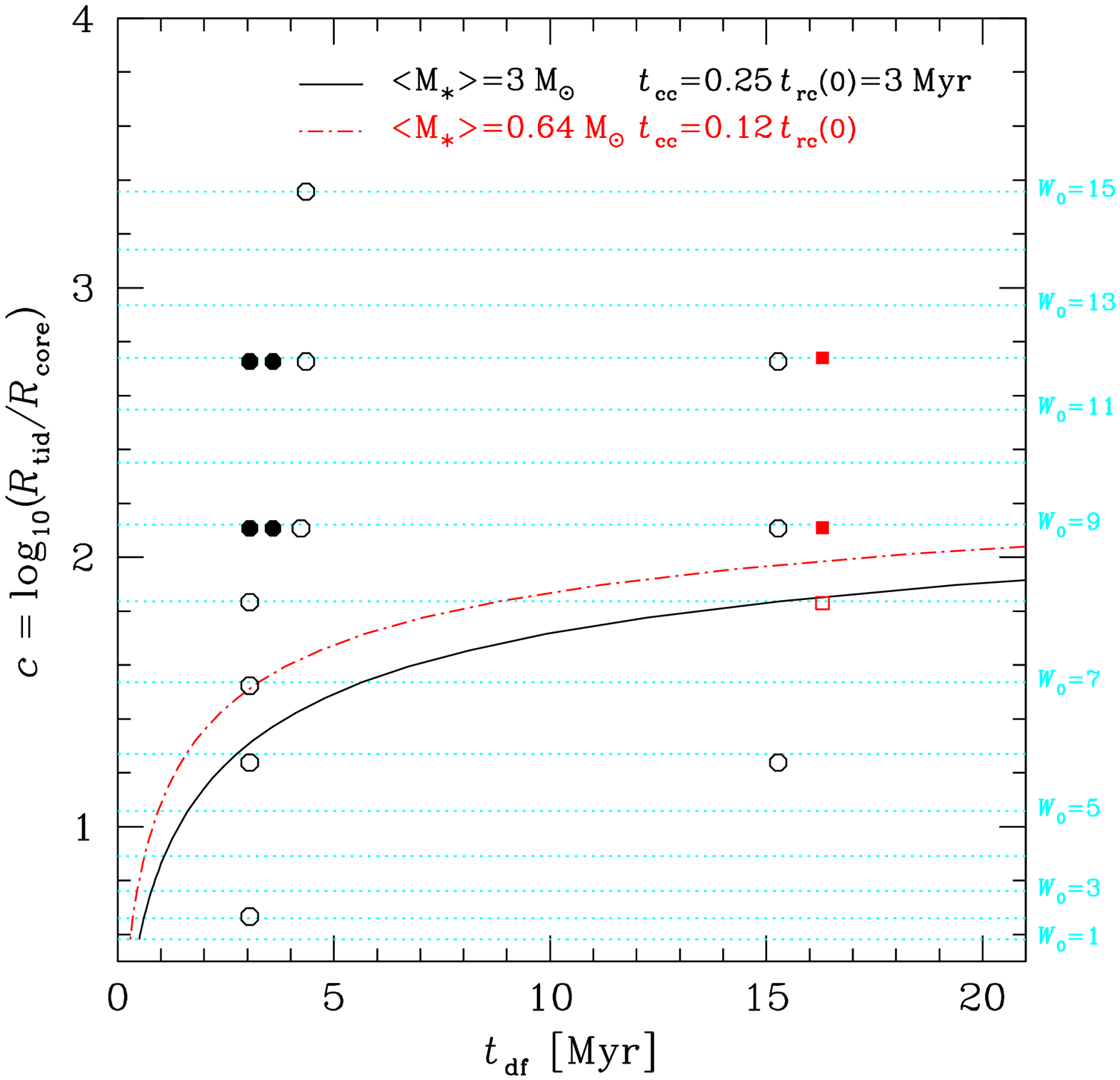}%
  }}
\caption{ 
Conditions for collisional runaway for models of the clusters MGG-9
and MGG-11 in M\,82 \citep{McCradyEtAl03}. This diagram is
adapted from Fig.~3 of \citet{PZBHMM04}. Their dynamical friction time
$t_{\rm df}$ is approximately equal to $4.7\times 10^{-3}\,T_{\rm FP}$
for models with $\langle M_\ast \rangle = 3\,\Msun$ and $1.3\times
10^{-3}\,T_{\rm FP}$ for $\langle M_\ast \rangle = 0.64\,\Msun$. $c$
is the initial concentration parameter; corresponding $W_0$ values are
given on the right of the diagram. Round points are for simulations by
\citeauthor{PZBHMM04} ($\langle M_\ast
\rangle = 3\,\Msun$), squares for our MGG-9 MC runs ($\langle M_\ast \rangle
= 0.64\,\Msun$). Solid symbols indicate that the model underwent
collisional runaway. The models on the left side (low $t_{\rm df}$
values) are for MGG-11, the ones on the right for MGG-9. The solid and
dash-dotted lines indicate where results of GFR04 predict
mass-segregation-driven core collapse to happen at $t=t_\ast=3$\,Myr
for the two different IMFs.}
\label{fig:PZetAl04}
\end{figure}
 
\citet{PZBHMM04}
have performed $N$-body simulations of the clusters MGG-9 and MGG-11
in M\,82 using $\sim 130\,000$ to $\sim 600\,000$ particles. Following
\citet{McCradyEtAl03}, they assume for MGG-9 an initial mass
of $1.8\times 10^6\,\Msun$ and a half-mass radius of $2.6$\,pc. Unlike
the case of MGG-11 for which there are indications of a lower cut-off
of the IMF at $\sim 1\,\Msun$ (which yields $\langle M_\ast
\rangle \simeq 3\,\Msun$), the observational data for MGG-9 is
compatible with a normal cluster Kroupa IMF extending from 0.1 to
100\,{\Msun} \citep{Kroupa00b}: $d{\Nstar}/dM_\ast \propto M_\ast^{-\alpha}$ with
$\alpha=1.3$ below $0.5\,\Msun$ and $\alpha=2.3$ for higher masses ($\langle M_\ast
\rangle \simeq 0.64\,\Msun$). To
get the total cluster mass as estimated observationally from its size
and velocity dispersion, some $2.7\times 10^6$ stars are
required. Because this number is too high for present-day direct
$N$-body simulations, \citet{PZBHMM04} adopted a $1\,\Msun$
cut-off. They considered King models with $W_0$ ranging from 6 to 12;
none of them led to collisional runaway, in contrast to their MGG-11
models. 

In light of GFR04, this is a surprising
result. From its position in Fig.~1 of Paper~I, the MGG-9
model is expected to undergo quick core collapse and runaway
collisions if its initial concentration corresponds to $W_0\gtrsim
8.5$. We have run three simulations of this cluster with
${\Npart}={\Nstar}=2.7\times 10^6$ for $W_0=8$, 9 and 12 (\Sim{MGG9-K8},
\Sim{MGG9-K9} and \Sim{MGG9-K12a}). The model with the lowest concentration
does not experience runaway but the other two do. We did not try to
simulate MGG-11. If the average stellar mass is as high as $3\,\Msun$,
the number of stars in that cluster is of order $\sim 10^5$, too small
for robust MC treatment with a broad IMF and $W_0\gtrsim3$. We note,
however that the condition for runaway suggested by GFR04,
$\tccrlx<t_\ast=3$\,Myr, predicts that many more of the MGG-11 models
considered by
\citeauthor{PZBHMM04} should have experienced collisional runaway (see
Fig.~\ref{fig:PZetAl04}). This disagreement indicates again that our
simple $\tccrlx$ criterion only applies to systems containing a
sufficiently large number of stars. In particular, for relatively low
${\Nstar}$, binaries can form through three-body processes and suspend
core collapse before collisions occur. Furthermore, systems of
very high concentration and (relatively) low number of stars, contain
a very low number of high-mass stars in their core, initially. Hence,
a clean mass-segregation-induced core collapse of the type found in
our MC simulations may not be possible.
 
\subsubsection{Loss-cone effects}
\label{subsec:loss_cone}

\begin{figure}
  \centerline{\resizebox{\hsize}{!}{
\includegraphics[bb=20 148 562  689,clip]{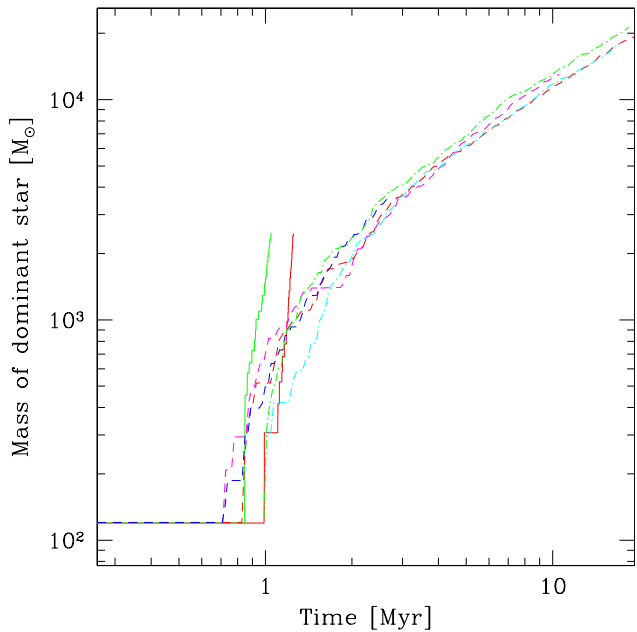}
 }} %
\caption{ 
Growth of the runaway star. We compare the results of simulations
using standard collision treatment (with sticky sphere approximation)
with runs for which the runaway star is considered as a fixed central
object and the loss-cone effects are taken into account following the
treatment of \citet{FB02b}. All runs are for a cluster with $W_0=8$,
${\Nstar}={\Npart}=1.45\times 10^6$ and ${\Rnb}=0.96$\,pc. The
curves in solid line are for the normal collision treatment
(\Sim{K8-14} and \Sim{K8-15}). The dashed curves are cases for which
a central fixed star of $120\,\Msun$ is present from the beginning
(\Sim{K8-11}, \Sim{K8-12}, \Sim{K8-17}). To obtain the two
dash-dotted curves, we introduced a central stellar object in a
cluster which had evolved to deep core collapse using normal collision
prescription (and no central object). The simulations with normal
collision treatment were interrupted during the runaway phase (models
\Sim{RA095} and \Sim{RA096}). Stellar evolution was not taken into
account except in one run with loss-cone treatment (\Sim{K8-17}), in
which case the VMS left the MS at $t=2.8$\,Myr when its mass had
reached $3700\,\Msun$.}
\label{fig:LC_comp}
\end{figure}

The MC treatment of collisions is based on the assumption that the two
neighbouring particles picked up at a given step as potential
collision partners are members of the same (local) distribution. In
this approach, one does not actually test whether the orbits of the two
particles would lead them to collide with each other. Instead, it is based on
the average collision probability between such stars in an imaginary
volume over which the properties of the stellar population (density,
mass spectrum, velocity distributions\ldots) should be homogeneous. For
this method to yield accurate collision rates, the cluster
properties must be relatively constant over the range of radii used
to estimate the local stellar density\footnote{In the innermost
parts of the cluster, this density is computed from the volume of
the spherical shell containing 17 nearby particles
\citep{Henon73}.}. A more subtle condition, in cases where one particle has 
unique properties and cannot be thought of as a typical member of
some continuous distribution, is that its orbit should allow it to
explore a significant fraction of this imaginary volume (for which the
stellar density and collisions times are computed) so that its
collision probability is also relatively homogeneous. 

Unfortunately, these conditions may cease to be fulfilled when the
runaway star has grown to such a mass that it basically stays at the
centre of the cluster with little radial motion in comparison with the
size of the orbits of nearby stars. The situation then becomes similar
to that of a massive (or intermediate-mass) black hole at a the centre
of a galactic nucleus (or globular cluster). While the VMS grows by
merging with stars coming into physical contact with it, the (I)MBH
destroys stars venturing into the Roche zone and accretes (some of)
their material. In both cases, stars must come very close to the
cluster centre, i.e., have very small pericentre distance $R_{\rm peri}$,
so the collision or disruption rate is dominated by stars on
high-eccentricity orbits. How these orbits are repopulated by 2-body
relaxation is known as the ``loss-cone replenishment problem''
\citep{FR76,LS77}. The basic difficulty is that relaxation may cause large
relative changes in $R_{\rm peri}$ in only a small fraction of the
relaxation time, i.e., the time over which the energy of the orbit
would be strongly affected. To give an illustration, for a Keplerian
orbit of eccentricity $e\approx 1$, The ``pericentre relaxation time''
is of order $t_{\rm r,peri}\simeq (1-e)\trlx$ where $\trlx$ is the
usual (energy) relaxation time. Let $P_{\rm orb}$ be the orbital
period of the star (from one pericentre passage to the next). For
regions of phase space where $P_{\rm orb} \ll t_{\rm r,peri}$, the
loss cone is empty, while in the opposite regime it is full (since
relaxation replenishes loss-cone orbits as quickly as stars are
removed through interactions with the central object). According to
standard loss-cone theory, in most situations, the disruption rate is
dominated by stars at the boundary between the two regimes. Hence, to
include loss-cone physics correctly into stellar dynamical models, one
should in principle resolve the effects of relaxation on timescales
much shorter than $\trlx$, actually as short as $P_{\rm orb}$ for
stars in the critical regime. The MC scheme used here does not allow
this directly because the time step $\delta t$ of a given particle can
only be a function of its present radial position $R$, not of its
eccentricity. Furthermore, $\delta t(R)$ must be an increasing
function of $R$ so that setting $\delta t=P_{\rm orb}$ at any given
$R$ would impose time steps at least as short as this for all
particles interior to $R$. To handle the loss-cone problem in a
satisfying way despite of this, an approximate treatment of the effect
of relaxation on timescales shorter than $\delta t$ has been devised;
it is described in detail in \citet{FB02b}. But this approach, as all
standard loss-cone theories, is based on the assumption that the
central object is rigorously fixed at $R=0$; in other terms, no
account is made of its ``wandering'' through Brownian motion.

In summary, our present code allows us to follow the growth of a VMS
using two opposite simplifying assumptions. (1) The ``standard
collisional treatment'' which computes collision probabilities with
the ``$nS_{\rm coll} \Vrel$'' formula assuming a small central homogeneous box,
ignoring
small-number effects and the small amplitude of the VMS motion. (2)
The ``loss-cone approach'' which neglects this motion altogether but
looks at the possibility of collision between the VMS and a given star
with (approximate) account of the specific orbit of the star and its
evolution (due to relaxation) on appropriately small timescales. A
limitation of the second approach, as implemented in the MC code is
that it only works for interactions with the central object leading to
disappearance of the star in one pericentre passage. In consequence,
here, we can only allow mergers but not fly-by collisions.  

In Fig.~\ref{fig:LC_comp}, a comparison is made of the results of both
treatments for a cluster with $W_0=8$, ${\Nstar}=N_{\rm p}=1.45\times
10^6$ and ${\Rnb}=0.96$\,pc. Obviously, after the first few
collisions, the growth of the VMS becomes much slower with the
loss-cone approach, probably as a result of loss-cone orbits being
rapidly depleted in the innermost region. In this situation, stellar
evolution has a chance to catch up with mergers and to terminate the
growth of the VMS as the interval of time between successive
collisions decreases. For this particular cluster model, when stellar
evolution is allowed, the runaway star still grows to more than
$3000\,\Msun$ before it leaves the MS (models \Sim{K8-16} and
\Sim{K8-17}). This happens around $t=3$\,Myr but, thanks to
collisional rejuvenation, the effective age of the VMS (its MS
lifetime) is only 1.4\,Myr. If the collision rate is so reduced after
the onset of runaway growth, clusters which experience core collapse
within just a little less than 3\,Myr can probably not form a VMS more
massive than a few hundred $\Msun$ before growth is interrupted by the
evolution of the VMS or normal high-mass stars off the MS.

The correct VMS growth rate, assuming the same $M$--$R$ relation and
collision physics, is very likely between what is found with our two
different treatments of the central object. Which approach is the most
accurate is difficult to tell. The effects of the wandering of the
central object on rates of interaction with stars remain to be
combined with the loss-cone theory
\citep[see][for short discussions]{Young77a,MT99,Sigurdsson03}. Considering 
length scales pertaining to the interactions of the VMS with the
cluster indicates that they take place in a complex regime. We take as
an example, simulation \Sim{K8-29} for which we have the best data
about the central conditions in the runaway phase. The MC simulation
indicates that the VMS has a wandering radius $\Rwand$ of a few
$10^{-4}$ (in $N$-body units or pc) by the time it has grown heavier
than $1000\,\Msun$. This value is in agreement with a simple analysis
based on the assumption of energy equipartition with the surrounding
stars, $M_{\rm VMS}V_{\rm wand}\approx \langle M_\ast
\rangle \sigmaOneD$, with $\langle M_\ast \rangle \approx 50\,\Msun$, and
equal amount of kinetic and potential energy in the cuspy density
background of these stars, $\rho(R)\propto R^{-1.5}$. However,
even though the motion of a massive object in a background of lighter
particles interacting with it gravitationally obeys energy
equipartition approximately \citep{CHL02,DHM03,LM04},
the energy dissipation introduced by collisions can lead to
non-equipartition, so the agreement may be fortuitous. In any case,
the wandering radius of the VMS is still much larger than its radius
($\sim 50\,\Rsun\simeq 10^{-6}$\,pc) so it cannot, in principle, be
considered strictly fixed at the centre. On the other hand, the
central number density (estimated inside $R=1.2\times 10^{-3}$ for 18
successive snapshots) is $n_0\simeq 1.5\times 10^8$, corresponding to
an average distance between stars of $\langle d \rangle = n_0^{-1/3}
\simeq 2\times 10^{-3}$ and the influence radius of the VMS is
formally $R_{\rm h}=GM_{\rm VMS}/(3\sigmaOneD^2)\approx 10^{-3}$
\citep{FR76}. These scales are significantly larger than $\Rwand$,
pointing to the breakdown of our standard collision treatment already
when $M_{\rm VMS}\simeq 1000\,\Msun$. The two approaches we have tried
make strong simplifications of a situation that can probably only be
treated correctly by direct orbital integration using, for instance,
some hybrid numerical code in which the central region of the cluster
is integrated with a direct $N$-body scheme, while the outer region is
modelled in the FP approximation.

\section{SUMMARY AND DISCUSSION}

\subsection{Summary of results}

This work is a continuation of our study of the dynamical evolution of
young, dense stellar clusters. Our goal is to determine
under which conditions an IMBH can form at the centre of such
systems. We are exploring the collisional runaway route. This scenario
applies to clusters in which the accumulation of massive main-sequence
stars ($\Mstar\simeq10-120\,\Msun$) at the centre, through the
relaxation-driven process of mass segregation, leads to the Spitzer
instability, thus triggering core collapse, before these stars evolve
to become compact objects, i.e., within $\sim 3\,$Myr. As the central
density of massive stars increases and their velocity dispersion
decreases, direct collisions between stars must occur, leading to the
possibility of successive mergers and the runaway formation of one
very massive star (VMS, $\Mstar\gtrsim 400\,\Msun$). Such a VMS is
a possible IMBH progenitor.

In GFR04, we have established the cluster conditions required for
the core collapse to be sufficiently rapid. In the present work, we
have investigated the next stage by implementing stellar
collisions. We have followed the core collapse and collisional runaway
using a Monte Carlo simulation code specially devised to account for
stellar collisions in a realistic way ({\MESSY}). In GFR04, 2-body relaxation
was the only physical process considered in most simulations so that,
as expected, up to statistical fluctuations, the results
 did not depend on the physical
size of the cluster or on the number of stars (provided there are
enough for the core to be resolved). The significant free
parameters were the initial structure of the cluster (profiles of
density, velocity and, in some cases, mass segregation) and the
IMF. Here, two more ingredients, collisions and stellar evolution, are
introduced, which break this degeneracy. For instance, the ratio of the
physical central velocity dispersion to the escape velocity from a star
(hence $\Nstar\Rstar/\Rh$) determines gravitational focusing and the
outcome of collisions and, more importantly, the ratio $\tcoll/\trlx$.

We have computed more than 100 models, varying the cluster size ($\Rh$
in the range $0.03-5$\,pc), mass ($\Nstar=10^5-10^8$) and
concentration. For the latter parameter, we have considered King
models with $W_0=3$ and $W_0=8$. For most simulations we have used a
Salpeter IMF extending from $0.2$ to $120\,\Msun$. We treated
collisions either in the sticky-sphere approximation or by making use
of prescriptions for merger conditions and mass loss derived from a
large set of SPH simulations. We used up to $9\times 10^6$ particles
(one simulation) but, because of the large number of simulations
and the relatively large amount of CPU time required to track the
runaway growth (requiring very short
time steps), we had to perform most high-$\Nstar$ runs using a number
of particles lower than $\Nstar$. Taking advantage of the statistical
nature of the MC algorithm, our code scales physical
processes so that each particle may represent a given (fixed) number
of stars, not necessarily one. This feature may legitimately be
questioned when a runaway happens as one expects only one star to
experience it. However, by varying the number of particles (but
keeping $\Nstar$ fixed) for a few cluster simulations, we have checked
that using $\Npart<\Nstar$ does not appear to produce spurious results
(such as, e.g., preventing the runaway).

We found that runway VMS formation occurs in {\em all cases\/} for which the
core collapse time is shorter than the MS lifetime
of massive stars, as predicted in GFR04. Furthermore, for very
massive clusters ($\gtrsim 10^7\,\Msun$), the core collapse is
facilitated (and sometimes driven) by collisions themselves, an effect
which extends the runaway domain to clusters of larger size than predicted by an
analysis based solely on relaxation. Such systems have
velocity dispersions in excess of $300\,\kms$ but we find that the
collisional mass loss and the reduced merger cross section cannot
prevent runaway, even at $\sim 1000\,\kms$.

We never observe the formation of more than one VMS. Only in clusters
that are initially collisional do collisions produce a high-mass tail
in the stellar mass spectrum. Most of the mass eventually incorporated
into the VMS originates in stars near the top of the IMF. Typically, for
clusters that are not initially collisional, $\gtrsim 90\,$\% of the
VMS mass is contributed by stars in the $60-120\,\Msun$ range. This is
due to the larger cross section of massive stars and, more
importantly, to the fact that collisions take place in the collapsed
core, which is dominated by these stars. In such clusters, most stars
that experience a collision do so only once, namely when they merge
with the growing VMS. In very dense clusters where collisions are
happening from the beginning of the evolution, the contribution of
lighter stars is slightly higher and a significant fraction of stars
that merge with the VMS have experienced at least one previous
collision.

Changing the parameters of the IMF modifies the condition for rapid core
collapse but has little qualitative effect on the runaway process
itself. In particular, clusters with a field Kroupa IMF \citep{KTG93},
which has an exponent $\alpha=2.7$ for massive stars, do experience
runaway in essentially the same way as those with a Salpeter IMF
($\alpha=2.35$), despite a much smaller total number of massive
stars. If the upper cutoff of the IMF is significantly lower than the
assumed $M_{\rm max}=120\,\Msun$, core collapse will take longer but
for $M_{\rm max}\lesssim 30\,\Msun$, this is over-compensated by a
longer MS lifetime. From one simulation we performed for $M_{\rm
max}=10\,\Msun$, we predict that the region of parameter space
leading to runaway is similar to what we found for
$M_{\rm max}=120\,\Msun$.

We found that in most circumstances, once the runaway has started,
the average time between successive mergers becomes much shorter than the
MS thermal relaxation timescale for the VMS, $\tKH$. Hence,
its structure and response to further collisions may be greatly
affected by its prior collision history. Modelling the complex
interplay between stellar dynamics, collisions, stellar structure and
stellar evolution which should eventually determine the final mass of
the VMS is presently not possible. Among the effects with potentially
strong bearing on the runaway process is the depletion of loss-cone
orbits that may come into play before the VMS has reached
$\sim1000\,\Msun$ as, being more massive than surrounding stars, it
becomes confined to a very small central region of the
cluster. Assuming a strictly fixed central object and using the
approximate treatment of loss-cone physics built into our MC code (for
the study of massive black holes in galactic nuclei), we have found a
significant reduction in the VMS growth rate (after the first $2-4$
mergers) but, in the few cases considered, the VMS still grew to a few
thousand $\Msun$ before evolving off the MS and the average time
between collisions, although $3-10$ times longer, remained below
$\tKH$ most of the time.

Our simulations allow us to establish the conditions for the onset of
runaway collisions and to study the early VMS growth. If globular
clusters are formed with a concentration equal or higher than a
$W_0=8$ King model (and if primordial binaries cannot prevent the runaway
process), a significant fraction ($\gtrsim 20$\,\%, see
Fig.~1 of Paper~I) may experience this phase in
the first few million years of their dynamical evolution.
Typically, a cluster containing $\sim10^6$ stars at this
concentration needs to have a half-mass radius smaller than $\Rh\simeq
2\,$pc to experience runaway, but a proto-galactic nucleus with $10^8$
stars has to be more compact than $\Rh\simeq 0.8\,$pc.

\subsection{Discussion}
\label{subsec:discu}

Many aspects of the runaway collision scenario and how we have
approached it in our numerical simulations have been discussed
already. In particular, the choice of physical ingredients considered
and the uncertainties and simplifications involved in their treatment
have been presented in Paper~I. In the results section of the present
paper, we have discussed other important uncertain aspects of the
process, namely the impact of collisions on the structure and growth
of the VMS and the possible effects of loss-cone depletion. In this last
subsection, we consider the role of interstellar gas.

\subsubsection{Role of residual gas in young clusters}
\label{subsec:residgas}
{
Because the runaway scenario presented here has to operate within the
first few Myr of the cluster life, it has, in fact, to be considered
in the framework of cluster formation, which takes place on a similar
timescale. No only does it mean, as mentioned above, that stars below
$2-3\,\Msun$ may still be on the pre-MS, with much larger radii than
adopted here, but, more importantly, that a significant amount of
residual gas may be present. Observations show that when a cluster
forms, not more than 30\,\% of the gas is eventually turned into stars
\citep{Lada99} but that clusters like R136 or NGC\,3603, as young as
$1-2$\,Myr, are already devoid of gas \citep[e.g.,][]{SBBZG04}. In
such clusters, the expulsion of gas (originally the dominant mass
component) is driven by the ionising radiation and winds of OB stars
and occurs impulsively. A large fraction of the stars are then lost
from the cluster and the remaining bound cluster swells quite
dramatically
\citep{KAH01,BK03a,BK03b}, an event likely to terminate the 
collisional phase after $\sim 1\,$Myr rather than $\sim 3$\,Myr. In
clusters with an escape velocity much higher than the typical
expansion velocity of the ionised gas, i.e., $\sim 10\,\kms$, complete
expulsion of the gas probably only occurs when the first SN explodes.

The faster evolution due to the much smaller size of the cluster in
the embedded phase counterbalances the possible reduction in the time
available for collisional runaway. Assuming that, when gas is
expelled, the mass of the cluster $M_{\rm cl}$ (including gas) is
reduced by a factor of $\sim 4$ and its size $\Rh$ expands $\sim 10$
times, the true initial relaxation time ($\trlx\propto
\sigmaOneD^3 n^{-1} \propto M_{\rm cl}^{3/2} \Rh^{3/2}$) should be of 
order 4 times shorter than what is inferred from observations of the
gas-free cluster. Furthermore, the collision time ($\tcoll \propto
\sigmaOneD n^{-1} \propto M_{\rm cl}^{-1/2} \Rh^{5/2}$) may have been
some 1000 times shorter, suggesting that a large fraction of clusters
may actually be collisional at birth.

The standard theory of star formation, which assumes spherically
symmetric accretion of gas on to the protostar, fails for stars more
massive than $10\,\Msun$ \citep{Yorke04,BZ05}. This problem led to the
proposal that the truly primordial IMF may not extend past
$10\,\Msun$ and that mergers may be ---at least partially--- responsible
for the formation of more massive stars
\citep{BBZ98,BB02,BZ05}. This would also naturally explain why these 
massive stars are preferentially found in the central regions of
clusters that are possibly too young to have experienced relaxational
mass segregation \citep{BD98,deGrijsEtAl02c}. To occur
on a sufficiently short timescale, collisions require central stellar
densities orders of magnitude higher than what is observed in those
young clusters but similar to what it may have been in the embedded
phase. 

Realistic simulations of embedded clusters should include the role of
gas, since its mass is likely to be more than twice that in
stars. Especially important are the increased velocity dispersion and
accretion from the ambient gas. The pioneering SPH/$N$-body
simulations of very early cluster evolution could only handle up to a
few hundred stars
\citep{BVB04,BB05}. Our MC results (run \Sim{K3-20} presented
in Section~\ref{subsec:otherIMFs}) suggest that, in much larger systems,
collisions would lead to the formation of just one (very) massive star
rather than a continuous spectrum spreading from $10\,\Msun$ to $\sim
100\,\Msun$. However, this result applies to clusters driven to core
collapse by mass segregation. When the cluster is initially
collisional, more stars should take part in collisions and a high-mass
spectrum may develop in a more orderly way (see
Fig.~\ref{fig:coll_hist_HiColl}). Also, the collisional phase may be
terminated before a runaway collision sequence starts through gas
expulsion by massive stars.}

\subsubsection{Gas retention in galactic nuclei}
\label{subsec:gasretention}
  
Our predictable result that evolution of the massive stars off the MS
terminates the evolution towards core collapse stems from our
assumption that all gas released by the stars is completely (and
instantaneously) lost from the cluster.  If the gas is not totally
expelled, there may be no core re-expansion. The core may instead
remain very dense. This opens the possibility for a longer-term
collisional evolution but now there will be plenty of stellar BHs
around that may destroy MS stars. 

Over a long timescale, it is possible that star formation will convert
the gas into new stars \citep{Sanders70b,QS90}. All this becomes
obviously quite intricate but would have to be included to treat
realistically proto-galactic nuclei. Nowadays, multi-phase
dynamical simulations, including interstellar gas (possibly in two
phases), stars and dark matter, and the matter and energy exchanges
between these components, are commonly performed to investigate the
formation and evolution of galaxies
\citep[e.g.,][]{MH96,ANSE03I,ANSE03II,SH03}. These works are based on the 
use of collisionless $N$-body codes coupled with SPH. Including
detailed gas dynamics in models of relaxing clusters is extremely
chalenging because the gas should possess a complex, non spherically
symmetric structure, evolving on timescales much shorter than $\trlx$
\citep{CM99b,WBP99,RFMW04,CNSdM05}.

Note that, in clusters with very high velocity dispersions, where
collisions may significantly contribute to gas production, some of it
will probably be retained. Indeed this can be seen in the extreme
situation of a completely disruptive collision (a rare occurrence; see
\citealt{FB05}). In this case, the total energy of the colliding stars
is their kinetic energy at (relatively) large separation minus the sum
of their individual binding energy (a positive quantity, by
definition). None of the stars can have a velocity higher than the
cluster's escape velocity $V_{\rm esc}$ and because the binding energy
is positive, the energy available per unit mass of stellar gas has
to be smaller than $1/2\,V_{\rm esc}^2$ so at least some gas must stay
in the cluster to conserve energy.

\section*{Acknowledgements}

We thank Holger Baumgardt, Melvyn Davies, John Fregeau, Pavel Kroupa,
Jamie Lombardi, Simon Portegies Zwart, Rainer Spurzem and Hans
Zinnecker for useful discussions. MF is indebted to Pau Amaro Seoane
for his extraordinary support during his stay in Heidelberg. The work
of MF was funded in part by the Sonder\-forschungs\-bereich (SFB) 439
`Galaxies in the Young Universe' (subproject A5) of the German Science
Foundation (DFG) at the University of Heidelberg. This work was also
supported by NASA ATP Grants NAG5-13236 and NNG04G176G, and NSF Grant
AST-0206276 to Northwestern University.


\label{lastpage}

\onecolumn

\begin{deluxetable}{lllllllll}
\tabletypesize{\small}
\tablecaption{Properties of Simulated Clusters
\label{table_clust}}
\tablewidth{0pt}
\tablehead{
\colhead{Name} & \colhead{$W_{0}$} & \colhead{$\Nstar$} & \colhead{$\Npart$} & \colhead{$\Rnb$} & \colhead{Coll.}   & \colhead{Peculiarities} & \colhead{$\tra$}   & \colhead{$\tra$} \\
               &                   &                    &                          & \colhead{(pc)}         &                   &                        & \colhead{($T_{\rm FP}$)} & \colhead{(Myr)} 
}
\startdata
\texttt{K3-1}     & $3$ & $10^{5}$          & $10^{5}$          & $  0.3$ & SS      & $1-120\,\Msun$                    & $4.38\times 10^{-2}$& 2.77          \\
\texttt{K3-2}     & $3$ & $10^{5}$          & $10^{5}$          & $  0.4$ & SS      & \nodata                           & $1.08\times 10^{-2}$& 2.25          \\
\texttt{K3-3}     & $3$ & $10^{5}$          & $10^{5}$          & $  0.4$ & SS      & $1-120\,\Msun$                    & \nodata             & \nodata       \\
\texttt{K3-4}     & $3$ & $10^{5}$          & $10^{5}$          & $  0.5$ & SS      & \nodata                           & \nodata             & \nodata       \\
\texttt{K3-5}     & $3$ & $10^{5}$          & $10^{5}$          & $  0.5$ & SS      & \nodata                           & $1.05\times 10^{-2}$& 3.03          \\
\texttt{K3-6}     & $3$ & $10^{5}$          & $10^{5}$          & $  0.5$ & SS      & \nodata                           & $1.07\times 10^{-2}$& 3.08          \\
\texttt{K3-7}     & $3$ & $10^{5}$          & $10^{5}$          & $  0.5$ & SS      & \nodata                           & $1.02\times 10^{-2}$& 2.94          \\
\texttt{K3-8}     & $3$ & $10^{5}$          & $10^{5}$          & $  0.5$ & SS      & $1-120\,\Msun$                    & \nodata             & \nodata       \\
\texttt{K3-9}     & $3$ & $3\times 10^{5}$  & $3\times 10^{5}$  & $ 0.03$ & SS      & \nodata                           & $9.89\times 10^{-3}$& 0.0634        \\
\texttt{K3-10}    & $3$ & $3\times 10^{5}$  & $3\times 10^{5}$  & $  0.1$ & SS      & \nodata                           & $1.06\times 10^{-2}$& 0.414         \\
\texttt{K3-11}    & $3$ & $3\times 10^{5}$  & $3\times 10^{5}$  & $  0.1$ & SPH     & \nodata                           & $1.04\times 10^{-2}$& 0.406         \\
\texttt{K3-12}    & $3$ & $3\times 10^{5}$  & $3\times 10^{5}$  & $  0.2$ & SS      & \nodata                           & $1.05\times 10^{-2}$& 1.16          \\
\texttt{K3-13}    & $3$ & $3\times 10^{5}$  & $3\times 10^{5}$  & $  0.3$ & SS      & \nodata                           & $9.60\times 10^{-3}$& 1.95          \\
\texttt{K3-14}    & $3$ & $3\times 10^{5}$  & $3\times 10^{5}$  & $  0.3$ & SS      & Constant 5\,\% mass loss          & $1.03\times 10^{-2}$& 2.09          \\
\texttt{K3-15}    & $3$ & $3\times 10^{5}$  & $3\times 10^{5}$  & $  0.3$ & SS      & Constant 10\,\% mass loss         & $1.03\times 10^{-2}$& 2.09          \\
\texttt{K3-16}    & $3$ & $3\times 10^{5}$  & $3\times 10^{5}$  & $  0.3$ & SS      & Constant 30\,\% mass loss         & \nodata             & \nodata       \\
\texttt{K3-17}    & $3$ & $3\times 10^{5}$  & $3\times 10^{5}$  & $  0.3$ & SS      & Fixed central object              & $1.01\times 10^{-2}$& 2.05          \\
\texttt{K3-18}    & $3$ & $3\times 10^{5}$  & $3\times 10^{5}$  & $  0.3$ & X       & 5\,\% mass loss                   & $9.35\times 10^{-3}$& 1.9           \\
\texttt{K3-19}    & $3$ & $3\times 10^{5}$  & $3\times 10^{5}$  & $  0.3$ & SS      & Constant $R_\ast$ for VMS         & $9.66\times 10^{-3}$& 1.96          \\
\texttt{K3-20}    & $3$ & $3\times 10^{5}$  & $3\times 10^{5}$  & $  0.3$ & SS      & $0.2-10\,\Msun$                   & $9.04\times 10^{-2}$& 19.9          \\
\texttt{K3-21}    & $3$ & $3\times 10^{5}$  & $3\times 10^{5}$  & $  0.3$ & SS      & Kroupa $0.08-120\,\Msun$          & $1.02\times 10^{-2}$& 2.54          \\
\texttt{K3-22}    & $3$ & $3\times 10^{5}$  & $3\times 10^{5}$  & $  0.3$ & SPH     & \nodata                           & $8.88\times 10^{-3}$& 1.8           \\
\texttt{K3-23}    & $3$ & $3\times 10^{5}$  & $3\times 10^{5}$  & $  0.4$ & SS      & \nodata                           & $9.50\times 10^{-3}$& 2.97          \\
\texttt{K3-24}    & $3$ & $3\times 10^{5}$  & $3\times 10^{5}$  & $  0.4$ & SPH     & \nodata                           & $9.28\times 10^{-3}$& 2.9           \\
\texttt{K3-25}    & $3$ & $3\times 10^{5}$  & $3\times 10^{5}$  & $  0.4$ & SS      & \nodata                           & $9.69\times 10^{-3}$& 3.02          \\
\texttt{K3-26}    & $3$ & $3\times 10^{5}$  & $3\times 10^{5}$  & $  0.4$ & SS      & \nodata                           & $1.02\times 10^{-2}$& 3.18          \\
\texttt{K3-27}    & $3$ & $3\times 10^{5}$  & $3\times 10^{5}$  & $  0.4$ & SS      & Check for orbit overlap           & $9.22\times 10^{-3}$& 2.85          \\
\texttt{K3-28}    & $3$ & $3\times 10^{5}$  & $3\times 10^{5}$  & $  0.5$ & SS      & \nodata                           & \nodata             & \nodata       \\
\texttt{K3-29}    & $3$ & $3\times 10^{5}$  & $3\times 10^{5}$  & $  0.5$ & SS      & $0.2-10\,\Msun$                   & \nodata             & \nodata       \\
\texttt{K3-30}    & $3$ & $10^{6}$          & $10^{6}$          & $  0.1$ & SS      & Kroupa $0.01-120\,\Msun$          & $3.50\times 10^{-3}$& 0.327         \\
\texttt{K3-31}    & $3$ & $10^{6}$          & $10^{6}$          & $  0.2$ & SS      & \nodata                           & $7.07\times 10^{-3}$& 1.87          \\
\texttt{K3-32}    & $3$ & $10^{6}$          & $10^{6}$          & $  0.2$ & SPH     & \nodata                           & $8.00\times 10^{-3}$& 1.39          \\
\texttt{K3-33}    & $3$ & $10^{6}$          & $10^{6}$          & $  0.3$ & SS      & \nodata                           & $7.94\times 10^{-3}$& 2.54          \\
\texttt{K3-34}    & $3$ & $10^{6}$          & $10^{6}$          & $  0.4$ & SS      & \nodata                           & \nodata             & \nodata       \\
\texttt{K3-35}    & $3$ & $3\times 10^{6}$  & $3\times 10^{5}$  & $  0.3$ & SS      & \nodata                           & \nodata             & \nodata       \\
\texttt{K3-36}    & $3$ & $3\times 10^{6}$  & $3\times 10^{6}$  & $  0.1$ & SS      & \nodata                           & $6.44\times 10^{-3}$& 0.617         \\
\texttt{K3-37}    & $3$ & $3\times 10^{6}$  & $3\times 10^{6}$  & $  0.2$ & SS      & \nodata                           & $6.84\times 10^{-3}$& 1.85          \\
\texttt{K3-38}    & $3$ & $3\times 10^{6}$  & $3\times 10^{6}$  & $  0.2$ & SPH     & \nodata                           & $7.25\times 10^{-3}$& 1.97          \\
\texttt{K3-39}    & $3$ & $3\times 10^{6}$  & $3\times 10^{6}$  & $  0.3$ & SS      & \nodata                           & \nodata             & \nodata       \\
\texttt{K3-40}    & $3$ & $10^{7}$          & $3\times 10^{5}$  & $  0.1$ & SS      & \nodata                           & $7.75\times 10^{-3}$& 1.21          \\
\texttt{K3-41}    & $3$ & $10^{7}$          & $3\times 10^{5}$  & $  0.1$ & SS      & \nodata                           & $7.53\times 10^{-3}$& 1.18          \\
\texttt{K3-42}    & $3$ & $10^{7}$          & $3\times 10^{5}$  & $  0.2$ & SS      & \nodata                           & $8.79\times 10^{-3}$& 3.89          \\
\texttt{K3-43}    & $3$ & $10^{7}$          & $3\times 10^{5}$  & $ 0.25$ & SS      & \nodata                           & \nodata             & \nodata       \\
\texttt{K3-44}    & $3$ & $10^{7}$          & $3\times 10^{5}$  & $  0.3$ & SS      & \nodata                           & \nodata             & \nodata       \\
\texttt{K3-45}    & $3$ & $10^{7}$          & $3\times 10^{5}$  & $  0.4$ & SS      & \nodata                           & \nodata             & \nodata       \\
\texttt{K3-46}    & $3$ & $10^{7}$          & $10^{6}$          & $  0.1$ & SPH     & \nodata                           & $6.41\times 10^{-3}$& 0.998         \\
\texttt{K3-47}    & $3$ & $10^{7}$          & $10^{6}$          & $  0.2$ & SS      & \nodata                           & $6.27\times 10^{-3}$& 2.76          \\
\texttt{K3-48}    & $3$ & $10^{7}$          & $10^{6}$          & $ 0.25$ & SS      & \nodata                           & \nodata             & \nodata       \\
\texttt{K3-49}    & $3$ & $3\times 10^{7}$  & $3\times 10^{5}$  & $  0.1$ & SS      & \nodata                           & $5.24\times 10^{-3}$& 1.3           \\
\texttt{K3-50}    & $3$ & $10^{8}$          & $3\times 10^{5}$  & $ 0.03$ & SS      & \nodata                           & $3.49\times 10^{-4}$& 0.0237        \\
\texttt{K3-51}    & $3$ & $10^{8}$          & $3\times 10^{5}$  & $ 0.04$ & SPH     & \nodata                           & $2.99\times 10^{-3}$& 0.312         \\
\texttt{K3-52}    & $3$ & $10^{8}$          & $3\times 10^{5}$  & $ 0.05$ & SS      & \nodata                           & $7.36\times 10^{-4}$& 0.107         \\
\texttt{K3-53}    & $3$ & $10^{8}$          & $3\times 10^{5}$  & $ 0.05$ & SS      & No relaxation                     & $8.90\times 10^{-4}$& 0.13          \\
\texttt{K3-54}    & $3$ & $10^{8}$          & $3\times 10^{5}$  & $  0.1$ & SS      & \nodata                           & $1.94\times 10^{-3}$& 0.801         \\
\texttt{K3-55}    & $3$ & $10^{8}$          & $3\times 10^{5}$  & $  0.2$ & SS      & \nodata                           & $3.41\times 10^{-3}$& 3.98          \\
\texttt{K3-56}    & $3$ & $10^{8}$          & $3\times 10^{5}$  & $  0.2$ & SPH     & \nodata                           & \nodata             & \nodata       \\
\texttt{K3-57}    & $3$ & $10^{8}$          & $3\times 10^{5}$  & $ 0.25$ & SS      & \nodata                           & \nodata             & \nodata       \\
\texttt{K3-58}    & $3$ & $10^{8}$          & $3\times 10^{5}$  & $  0.3$ & SS      & \nodata                           & \nodata             & \nodata       \\
\texttt{K3-59}    & $3$ & $10^{8}$          & $3\times 10^{5}$  & $  0.4$ & SS      & \nodata                           & \nodata             & \nodata       \\
\texttt{K3-60}    & $3$ & $10^{8}$          & $3\times 10^{5}$  & $  0.5$ & SS      & \nodata                           & \nodata             & \nodata       \\
\texttt{K3-61}    & $3$ & $10^{8}$          & $10^{6}$          & $ 0.04$ & SS      & \nodata                           & $5.05\times 10^{-4}$& 0.0524        \\
\texttt{K3-62}    & $3$ & $10^{8}$          & $10^{6}$          & $  0.1$ & SPH     & \nodata                           & $3.19\times 10^{-3}$& 1.31          \\
\texttt{K3-63}    & $3$ & $10^{8}$          & $10^{6}$          & $  0.1$ & SPH     & \nodata                           & $3.20\times 10^{-3}$& 1.31          \\
\texttt{K3-64}    & $3$ & $10^{8}$          & $10^{6}$          & $  0.1$ & SS      & \nodata                           & $1.90\times 10^{-3}$& 0.78          \\
\texttt{K3-65}    & $3$ & $10^{8}$          & $10^{6}$          & $  0.2$ & SPH     & \nodata                           & \nodata             & \nodata       \\
\texttt{K8-1}     & $8$ & $3\times 10^{5}$  & $3\times 10^{5}$  & $  0.9$ & SS      & \nodata                           & $3.97\times 10^{-4}$& 0.418         \\
\texttt{K8-2}     & $8$ & $3\times 10^{5}$  & $3\times 10^{5}$  & $    2$ & SS      & \nodata                           & $4.63\times 10^{-4}$& 1.62          \\
\texttt{K8-3}     & $8$ & $3\times 10^{5}$  & $3\times 10^{5}$  & $    3$ & SS      & \nodata                           & \nodata             & \nodata       \\
\texttt{K8-4}     & $8$ & $3\times 10^{5}$  & $3\times 10^{5}$  & $    4$ & SS      & \nodata                           & \nodata             & \nodata       \\
\texttt{K8-5}     & $8$ & $3\times 10^{5}$  & $3\times 10^{6}$  & $  1.2$ & SPH     & \nodata                           & $3.89\times 10^{-4}$& 0.618         \\
\texttt{K8-6}     & $8$ & $3\times 10^{5}$  & $3\times 10^{6}$  & $  4.7$ & SPH     & \nodata                           & \nodata             & \nodata       \\
\texttt{K8-7}     & $8$ & $10^{6}$          & $10^{6}$          & $  0.1$ & SS      & \nodata                           & $4.00\times 10^{-4}$& 0.0246        \\
\texttt{K8-8}     & $8$ & $10^{6}$          & $10^{6}$          & $    1$ & SS      & \nodata                           & $4.32\times 10^{-4}$& 0.841         \\
\texttt{K8-9}     & $8$ & $10^{6}$          & $10^{6}$          & $    2$ & SS      & \nodata                           & $4.82\times 10^{-4}$& 2.65          \\
\texttt{K8-10}    & $8$ & $10^{6}$          & $10^{6}$          & $    3$ & SS      & \nodata                           & \nodata             & \nodata       \\
\texttt{K8-11}    & $8$ & $1.4\times 10^{6}$& $1.4\times 10^{6}$& $ 0.96$ & SS      & Fixed central object              & $4.02\times 10^{-4}$& 0.843         \\
\texttt{K8-12}    & $8$ & $1.4\times 10^{6}$& $1.4\times 10^{6}$& $ 0.96$ & SS      & Fixed central object              & $3.91\times 10^{-4}$& 0.819         \\
\texttt{K8-13}    & $8$ & $1.4\times 10^{6}$& $1.4\times 10^{6}$& $ 0.96$ & SS      & Fixed central object, tid. trunc. & $3.75\times 10^{-4}$& 0.786         \\
\texttt{K8-14}    & $8$ & $1.4\times 10^{6}$& $1.4\times 10^{6}$& $ 0.96$ & SS      & \nodata                           & $4.66\times 10^{-4}$& 0.976         \\
\texttt{K8-15}    & $8$ & $1.4\times 10^{6}$& $1.4\times 10^{6}$& $ 0.96$ & SS      & \nodata                           & $3.98\times 10^{-4}$& 0.834         \\
\texttt{K8-16}    & $8$ & $1.4\times 10^{6}$& $1.4\times 10^{6}$& $ 0.96$ & SS      & Fixed central object, stell. evol.& $4.19\times 10^{-4}$& 0.878         \\
\texttt{K8-17}    & $8$ & $1.4\times 10^{6}$& $1.4\times 10^{6}$& $ 0.96$ & SS      & Fixed central object, stell. evol.& $3.91\times 10^{-4}$& 0.819         \\
\texttt{K8-18}    & $8$ & $1.4\times 10^{6}$& $1.4\times 10^{6}$& $ 0.96$ & SS      & \nodata                           & $4.32\times 10^{-4}$& 0.904         \\
\texttt{K8-19}    & $8$ & $3\times 10^{6}$  & $3\times 10^{6}$  & $  0.2$ & SPH     & \nodata                           & $3.31\times 10^{-4}$& 0.0892        \\
\texttt{K8-20}    & $8$ & $3\times 10^{6}$  & $3\times 10^{6}$  & $    1$ & SS      & \nodata                           & $3.43\times 10^{-4}$& 1.03          \\
\texttt{K8-21}    & $8$ & $3\times 10^{6}$  & $3\times 10^{6}$  & $    1$ & SPH     & \nodata                           & $3.43\times 10^{-4}$& 1.03          \\
\texttt{K8-22}    & $8$ & $3\times 10^{6}$  & $3\times 10^{6}$  & $    1$ & SS      & \nodata                           & $3.70\times 10^{-4}$& 1.12          \\
\texttt{K8-23}    & $8$ & $3\times 10^{6}$  & $3\times 10^{6}$  & $    1$ & SS      & \nodata                           & $3.74\times 10^{-4}$& 1.13          \\
\texttt{K8-24}    & $8$ & $3\times 10^{6}$  & $3\times 10^{6}$  & $    1$ & SS      & Fixed central object              & $2.99\times 10^{-4}$& 0.9           \\
\texttt{K8-25}    & $8$ & $3\times 10^{6}$  & $3\times 10^{6}$  & $  1.2$ & SPH     & \nodata                           & $4.22\times 10^{-4}$& 1.65          \\
\texttt{K8-26}    & $8$ & $3\times 10^{6}$  & $3\times 10^{6}$  & $  1.4$ & SS      & \nodata                           & $4.27\times 10^{-4}$& 2.13          \\
\texttt{K8-27}    & $8$ & $3\times 10^{6}$  & $3\times 10^{6}$  & $    2$ & SS      & \nodata                           & \nodata             & \nodata       \\
\texttt{K8-28}    & $8$ & $3\times 10^{6}$  & $3\times 10^{6}$  & $  4.7$ & SPH     & \nodata                           & \nodata             & \nodata       \\
\texttt{K8-29}    & $8$ & $9\times 10^{6}$  & $9\times 10^{6}$  & $    1$ & SS      & \nodata                           & $3.78\times 10^{-4}$& 1.78          \\
\texttt{K8-30}    & $8$ & $10^{7}$          & $3\times 10^{6}$  & $    1$ & SS      & \nodata                           & $3.61\times 10^{-4}$& 1.78          \\
\texttt{K8-31}    & $8$ & $10^{7}$          & $3\times 10^{6}$  & $  1.4$ & SS      & \nodata                           & $3.12\times 10^{-4}$& 2.55          \\
\texttt{K8-32}    & $8$ & $10^{7}$          & $3\times 10^{6}$  & $  1.4$ & SS      & \nodata                           & $3.86\times 10^{-4}$& 3.15          \\
\texttt{K8-33}    & $8$ & $10^{7}$          & $3\times 10^{6}$  & $    2$ & SS      & \nodata                           & \nodata             & \nodata       \\
\texttt{K8-34}    & $8$ & $3\times 10^{7}$  & $3\times 10^{6}$  & $    1$ & SS      & \nodata                           & $3.14\times 10^{-4}$& 2.45          \\
\texttt{K8-35}    & $8$ & $3\times 10^{7}$  & $3\times 10^{6}$  & $  1.4$ & SS      & \nodata                           & \nodata             & \nodata       \\
\texttt{K8-36}    & $8$ & $10^{8}$          & $3\times 10^{6}$  & $  0.1$ & SS      & \nodata                           & $7.49\times 10^{-5}$& 0.0308        \\
\texttt{K8-37}    & $8$ & $10^{8}$          & $3\times 10^{6}$  & $  0.6$ & SS      & \nodata                           & $2.40\times 10^{-4}$& 1.45          \\
\texttt{K8-38}    & $8$ & $10^{8}$          & $3\times 10^{6}$  & $  0.8$ & SS      & \nodata                           & $2.74\times 10^{-4}$& 2.55          \\
\texttt{K8-39}    & $8$ & $10^{8}$          & $3\times 10^{6}$  & $    1$ & SS      & \nodata                           & \nodata             & \nodata       \\
\texttt{K8-40}    & $8$ & $10^{8}$          & $3\times 10^{6}$  & $  1.4$ & SS      & \nodata                           & \nodata             & \nodata       \\
\texttt{K8-41}    & $8$ & $10^{8}$          & $3\times 10^{6}$  & $    2$ & SS      & \nodata                           & \nodata             & \nodata       \\
\texttt{MGG9-K8}  & $8$ & $2.7\times 10^{6}$& $2.7\times 10^{6}$& $  2.6$ & SPH     & Model for MGG-9                   & \nodata             & \nodata       \\
\texttt{MGG9-K9}  & $9$ & $2.7\times 10^{6}$& $2.7\times 10^{6}$& $  2.6$ & SPH     & Model for MGG-9                   & $1.55\times 10^{-4}$& 1.95          \\
\texttt{MGG9-K12a}& $12$& $2.7\times 10^{6}$& $2.7\times 10^{6}$& $  2.6$ & SPH     & Model for MGG-9                   & $5.10\times 10^{-6}$& 0.0644        \\
\texttt{MGG9-K12b}& $12$& $2.7\times 10^{6}$& $2.7\times 10^{6}$& $  2.6$ & SPH     & Model for MGG-9                   & $5.98\times 10^{-6}$& 0.0755        \\
\enddata

\tablecomments{ When not otherwise mentioned in the "Peculiarities"
column, clusters were modeled with a Salpeter ($\alpha=2.35$) IMF
extending from 0.2 to $120\,\Msun$, our standard $M$--$R$ relation and
prescription for collisional rejuvenation and MS lifetime. The data
listed in the column entitled "Coll." indicates the type of collision
treatment used. "SS" stands for sticky-sphere approximation; "SPH" for
the prescriptions for mass loss and merger derived from the SPH
simulations of Freitag \& Benz (2005); "X" is a special case for which a simple
merger criterion was used ($\lambda_{\rm
merg}=-1.14-0.5(l_v-3.0)$,
see Section 3.2.5 of Paper~I) and a flat mass-loss rate of 5\,\% was
assumed in all cases. $\tra$ is the time at which runaway growth of a
VMS started. It is given in Fokker-Planck time units ($T_{\rm FP}$)
and Myr. No value is given for clusters which didn't experience
collisional runaway.}

\label{table:simul_list} 
\end{deluxetable}

\end{document}